\documentclass[]{aa}

\usepackage[T1]{fontenc}
\usepackage{newtxtext,newtxmath}
\usepackage{graphicx}	
\usepackage{amsmath}	
\usepackage{tabularx}
\usepackage{placeins}

\usepackage{natbib,twoopt}
\usepackage[breaklinks=true]{hyperref} 
\bibpunct{(}{)}{;}{a}{}{,}             
\makeatletter
  \newcommandtwoopt{\citeads}[3][][]{\href{http://adsabs.harvard.edu/abs/#3}%
    {\def\hyper@linkstart##1##2{}%
     \let\hyper@linkend\@empty\citealp[#1][#2]{#3}}}
  \newcommandtwoopt{\citepads}[3][][]{\href{http://adsabs.harvard.edu/abs/#3}%
    {\def\hyper@linkstart##1##2{}%
     \let\hyper@linkend\@empty\citep[#1][#2]{#3}}}
  \newcommandtwoopt{\citetads}[3][][]{\href{http://adsabs.harvard.edu/abs/#3}%
    {\def\hyper@linkstart##1##2{}%
     \let\hyper@linkend\@empty\citet[#1][#2]{#3}}}
  \newcommandtwoopt{\citeyearads}[3][][]%
    {\href{http://adsabs.harvard.edu/abs/#3}
    {\def\hyper@linkstart##1##2{}%
     \let\hyper@linkend\@empty\citeyear[#1][#2]{#3}}}
\makeatother

\graphicspath{{./}{figures/}}

\defcitealias{seitenzahl:12}{S13\_DDT\_M1.40}
\defcitealias{leung2018}{L18\_DDT\_M1.3}
\defcitealias{townsley:16}{T16\_DDT\_M1.40}
\defcitealias{fink:2014}{F14\_PDf\_M1.40}
\defcitealias{kromer:15}{K15\_PDf\_M1.40}
\defcitealias{iwamoto:99}{I99\_PDf\_M1.40}
\defcitealias{pakmor:12}{P12\_PDt\_VM}
\defcitealias{sim:2010}{S10\_PDt\_M1.06}
\defcitealias{miles:19}{M19\_PDt\_M0.80}
\defcitealias{gronow:21gce}{G21\_DD\_M10\_03}
\defcitealias{leung:20}{L20\_DD}
\defcitealias{shen:2018}{S18\_DD\_M10\_00}
\defcitealias{limongi2018}{L\&C18}
\defcitealias{nomoto13}{N13}
\defcitealias{APOGEE:R13}{APOGEE}

\newcolumntype{P}[1]{>{\centering\arraybackslash}p{#1}}

\begin{document} 

\title{Using chemical evolution models of the Milky Way disk to constrain Type Ia supernova progenitors}

\titlerunning{Galactic chemical evolution with supernova}

   \author{T. C. L. Trueman
          \inst{1,2,3,4}
          \thanks{thomas.trueman@csfk.org}
          \and
          M. Pignatari
          \inst{1,2,3,4,5}
          \and
          B. Cseh
          \inst{1,2,7}
          \and
          J. D. Keegans
          \inst{3,4,5}
          \and
          B. Côté
          \inst{4,5,6}
          \and
          B. K. Gibson
          \inst{3,5}}

    \institute{Konkoly Observatory, HUN-REN Research Centre for Astronomy and Earth Sciences, H-1121 Budapest, Hungary
             \and
    CSFK, MTA Centre of Excellence, Budapest, Konkoly Thege Miklós út 15-17., H-1121, Hungary
    \and
    E.A. Milne Centre for Astrophysics, University of Hull, HU6 7RX, UK
        \and
    NuGrid Collaboration \thanks{http://nugridstars.org}
        \and
    Joint Institute for Nuclear Astrophysics—Center for the Evolution of the Elements (JINA-CEE), USA
        \and
    Department of Physics and Astronomy, University of Victoria, Victoria, BC V8P 5C2, Canada
        \and
    MTA-ELTE Lend{\"u}let "Momentum" Milky Way Research Group, Hungary
    }

   





    \abstract
    {Thermonuclear explosions of carbon-oxygen white dwarfs as Type Ia supernovae (SNe Ia) play a significant role in the galactic chemical evolution (GCE) of the Milky Way. However, a long-standing and as yet unresolved problem of modern astrophysics concerns the identity of their progenitor.}
    {We aim to use GCE predictions to help constrain potential SN Ia progenitor scenarios, since it is well known that SN Ia nucleosynthesis yields, in particular the Fe-peak elements, depend on the explosion mechanism.}
    {We calculated 1140 GCE models and compared the GCE-predicted abundance ratios for four different SN Ia explosion mechanisms -- two from sub-Chandrasekhar ($M_{\text{Ch}}$) mass progenitors and two from near-$M_{\text{Ch}}$ mass progenitors -- to spectroscopic measurements of Milky Way disk stars, considering both local thermodynamic equilibrium (LTE) and non-LTE (NLTE) assumptions.
    We calibrated the GCE framework for two sets of massive star yields in order to assess how stellar modelling uncertainties affect the relative contribution from core-collapse supernovae (CCSNe) towards Si, Ca, and the Fe-peak elements.}
    {From a GCE perspective, Si and Ca cannot be used to constrain SN Ia progenitors since there is little variation in their yields between different explosion types. The GCE of [Ti/Fe] and [Co/Fe] are not reproduced by any combination of yields. The [Cr/Fe] ratio is also problematic, since hardly any NLTE data of the disk are available and there are conflicting yields from CCSNe. For [Mn/Fe], neither CCSN yield set are compatible with the NLTE data. For [Ni/Fe], the NLTE data are well fit by one set of CCSN yields, with the best-fitting GCE models having a $\sim85\%$ contribution from sub-$M_{\text{Ch}}$ SNe Ia.}
    {We advise caution when using GCE models to constrain the Galaxy's SN Ia population, since the results depend on both the choice of CCSN yields and the elemental ratio used as a diagnostic.}


   \maketitle

%


\section{Introduction} \label{sec:intro}

Type Ia supernovae (SNe Ia) are some of the most luminous events in the universe. They are also appreciated as important beacons of knowledge in the fields of both galactic evolution and cosmology. On a galactic scale, they play a pivotal role in the chemical evolution of Fe and several of its neighbouring elements (Fe-peak elements) in the interstellar medium \citep[ISM;][]{matteucci:2021}. On a cosmological scale, SNe Ia can be used as standard candles for measuring distances due to the correlation between their peak luminosity and light-curve width \citep[the Phillips relation;][]{phillips:93}, and they were also credited with helping in the discovery of dark energy and the associated expansion of the Universe \citep{riess:1998, perlmutter:1999}. The fact that the identity of the progenitor and the explosion mechanism for SNe Ia remains unresolved has consequences for these observations \footnote{Sometimes aptly named the `progenitor problem' \citep[see, e.g.][]{maoz_mannucci_2012}}.

Some observational clues regarding the origin of SNe Ia are the absence of hydrogen and helium absorption lines in their spectra, as well as the production of large amounts $(\sim0.6M_{\odot})$ of radioactive $^{56}$Ni in their central regions; the decay of which is responsible for the sharp maxima in their light curves. The discovery of various sub-classes of SNe Ia that deviate spectroscopically from the characteristic `normal' explosions that follow the Phillips relation have fuelled the notion that more than one progenitor channel could lead to SNe Ia \citep[see, e.g.][]{tauben:2017}. Furthermore, the fact that SNe Ia are observed in both galaxies undergoing active star formation as well as quiescent lenticular and elliptical galaxies suggests the existence of two distinct populations of SN Ia progenitor that explode over different timescales \citep{scannapieco:05, man:2006}. 

There is a consensus at least that SNe Ia originate in binary systems in which a carbon-oxygen white dwarf (WD) accretes material from its stellar companion. As the WD grows in mass, its central temperature rises due to compressional heating, eventually triggering a thermonuclear explosion. A number of potential explosion mechanisms have been proposed, depending on whether material is donated from a non-degenerate star \citep[single-degenerate scenario;][]{whelan:1973} or another WD \citep[double-degenerate scenario;][]{iben:1984}, and whether the mass of the exploding WD is close to or below the Chandrasekhar mass $(M_{\text{Ch}}\sim1.4M_{\odot})$. 

The relative frequency of SNe Ia events in a population of stars from the single-degenerate and double-degenerate scenarios can be estimated with binary population synthesis (BPS) codes \citep{Belczynski:2008, Ruiter:2011, Temmink:2020}, although simplifying assumptions regarding the treatment of otherwise computationally intensive physical processes can lead to substantial differences in the predicted SN Ia rates for different BPS codes \citep{toonen:2014}. There are valid arguments favouring the double-degenerate scenario and the sub-$M_{\text{Ch}}$ SN Ia explosions from the theoretical point of view, due to the difficulty in forming a $M_{\text{Ch}}$ progenitor within the single-degenerate scenario \citep[e.g.][]{Denissenkov:2017, Battino:2020} and the lack of ionising UV radiation in extra-galactic observations \citep[e.g.][]{Woods:2013, Johansson:2016}. However, at present, it is unclear which binary configuration(s) may contribute to the SNe Ia that are observed in nature \citep[see, e.g. discussions in][]{hillebrandt:2013}. 

Given the uncertainty regarding both the physical progenitor system and the explosion mechanism, a wealth of SN Ia yields have been calculated using detailed stellar nucleosynthesis models covering a large parameter space of initial conditions \citep[for recent reviews regarding these different progenitor conditions, we refer the reader to][]{mario:2018, Soker:2019, Ruiter:2021}. Whilst the majority of SNe Ia have little variation in their spectroscopic and photometric features, their yields are chemically diverse and are particularly sensitive to both the density (mass) of the WD and the explosion mechanism. The yields are also metallicity dependent, as the metallicity of the zero age main sequence progenitor star determines the electron fraction ($Y_e$) during nuclear statistical equilibrium (NSE) in near-$M_{\text{Ch}}$ WDs, and the neutron excess during $\alpha$-rich freeze-out in sub-$M_{\text{Ch}}$ WDs \citep{Hartman:1985, gronow:21gce}. These nucleosynthesis yields are the cornerstone of galactic chemical evolution (GCE) models that aim to reproduce stellar abundances derived from spectroscopy. 

The W7 model of \cite{iwamoto:99} has historically been the SN Ia yield of choice for GCE models of the solar neighbourhood, due to its close fit to SN Ia spectra and light curves, as well as the fact it provides a reasonable fit to the elemental ratios measured in stars. These ratios include the increase in [Mn/Fe] and the decrease in [$\alpha$/Fe] after [Fe/H]$\;\sim-1$, which is a consequence of the delayed enrichment from SNe Ia in the ISM with respect to that of massive stars and core-collapse supernovae \citep[CCSNe;][]{minchev2013, snaith:2015}. However, the yields for the W7 model are indicative only of a 1D pure deflagration explosion in a near-$M_{\text{Ch}}$ WD, and the reaction network includes outdated electron capture rates that leads to over-estimations of [Ni/Fe] relative to the solar value. The deflagration of a near-$M_{\text{Ch}}$ WD is now considered the favoured explosion mechanism of the SN Iax sub-class of SN Ia that make up only a small percentage of the total SN Ia rate \citep{Foley:2013}.

In regards to `normal' SNe Ia, a large number of GCE studies have looked to constrain the relative fraction of SNe Ia that have either a sub- or near-$M_{\text{Ch}}$ progenitor \citep[see, e.g.][]{seitenzahl:Mn, kobayashi:mn, Delosreyes:2020, eitner:2020, palla:2021, Sanders:2021}. These studies commonly use Mn, or Ni, or both as tracer elements for WD mass, since they are produced during NSE at high densities \citep{Seitenzahl:2017, lach:2020, Keegans:2023}; this condition is only met in near-$M_{\text{Ch}}$ WD, therefore at least some fraction of the SNe Ia that exploded in the past must have been of near-$M_{\text{Ch}}$ origin in order to reach solar [Mn/Fe] and [Ni/Fe] \citep{seitenzahl:Mn, kobayashi:mn}. However, the sub/near progenitor ratio is far from settled. In particular, GCE models predict different estimates for the required fraction of sub-$M_{\text{Ch}}$ progenitor depending on the specific mixing and fallback parameters of the core-collapse supernovae yields \citep[see, e.g.][]{gronow:21gce}. Recently this issue was addressed by \cite{eitner:2020}, who applied a scaling factor to a set of CCSNe yields in order to account for the underestimation of [Mn/Fe] in the GCE models at lower metallicities when compared to non-local thermodynamic equilibrium (NLTE) observations. 

In this work we compare GCE predicted abundances of the Fe-peak elements (Ti, Cr, Co, Mn, and Ni) to observational measurements in stars in order to constrain SN Ia explosion mechanisms in the Galaxy. The GCE framework is calibrated for two sets of massive star yields, one including the effects of rotation, and GCE calculations are made for several SN Ia yields covering a range of potential explosion mechanisms. We are careful to discriminate among the observational datasets based on whether local thermodynamic equilibrium (LTE) or non-LTE (NLTE) is assumed, such that LTE and NLTE analyses are considered separately. Since no NLTE results exist for V, this element is omitted entirely from the analyses. We extend the scope of this investigation to elements other than the commonly used SN Ia progenitor tracers Mn and Ni, because (i) recently it has been shown that Cr and V could potentially be used to probe the explosion pattern of sub-$M_{\text{Ch}}$ SNe Ia \citep{leung:20, palla:2021} and (ii) the Fe-peak elements are made together in stellar ejecta and therefore their production will be connected. Furthermore, we consider also the evolution of the intermediate mass $\alpha$-elements Si and Ca, for which a non-negligible contribution from SNe Ia is required to fit the solar abundances \citep{seitenzahl:12, Kobayashi:2020a}. 

The paper is arranged as follows. In Section \ref{sec:yields} we describe the SN Ia yields that are investigated in this work, followed by a description of the GCE code in Section \ref{sec:GCE}. In Section \ref{sec:results} we compare the predicted GCE of the Fe-peak elements for each of the SN Ia yields to observational measurements in stars, we discuss also the differences between the GCE models with rotating and non-rotating massive star yields. GCE models with a combination of SN Ia yields are considered in Section \ref{sec:combos}, and a statistical test is used to determine the respective fraction of SNe Ia from the sub- and near-$M_{\text{Ch}}$ progenitor channels that best fits the data. Finally, in Section \ref{sec:conclusion} we offer a discussion of the results and our concluding remarks.

\section{The SN Ia yields}\label{sec:yields}

We investigate the GCE of SNe Ia in the Milky Way by considering the yields for nucleosynthesis models which cover a range of progenitor masses and explosion mechanisms.  For deflagration-to-detonation transition (DDT) explosions of near-$M_{\text{Ch}}$ progenitor we include the yields from \cite{seitenzahl:12}, \cite{leung2018}, and \cite{townsley:16}. The yields for pure deflagration (PDf) models are taken from \cite{iwamoto:99}, \cite{kromer:15} and \cite{fink:2014}. In terms of sub-$M_{\text{Ch}}$ progenitor, pure detonation (PDt) yields are taken from the models of \cite{pakmor:12} and \cite{sim:2010}, while we use the double-detonation (DD) yields of \cite{gronow:21gce}, \cite{leung:20}, and \cite{shen:2018}. We note that the above selection includes a mix of metallicity-dependent and single-metallicity SN Ia yields, where these last are calculated assuming an initial progenitor metallicity of $Z=Z_{\odot}$. Below we give a brief description of the main characteristics for each of the SN Ia models. Hereafter, the models are labelled according to the following naming convention: $(1)$ shorthand reference, $(2)$ explosion mechanism, $(3)$ progenitor mass in $M_{\odot}$, $(4)$ He shell mass in $M_{\odot}$ (DD models only), with $(1)$, $(2)$, $(3)$ and, where appropriate, $(4)$ all separated by an underscore. As an example, the $1.40M_{\odot}$, deflagration-to-detonation transition model of \cite{townsley:16} is labelled \citetalias{townsley:16}.

\begin{figure*}[!h]
    \centering
    \includegraphics[scale=0.4]{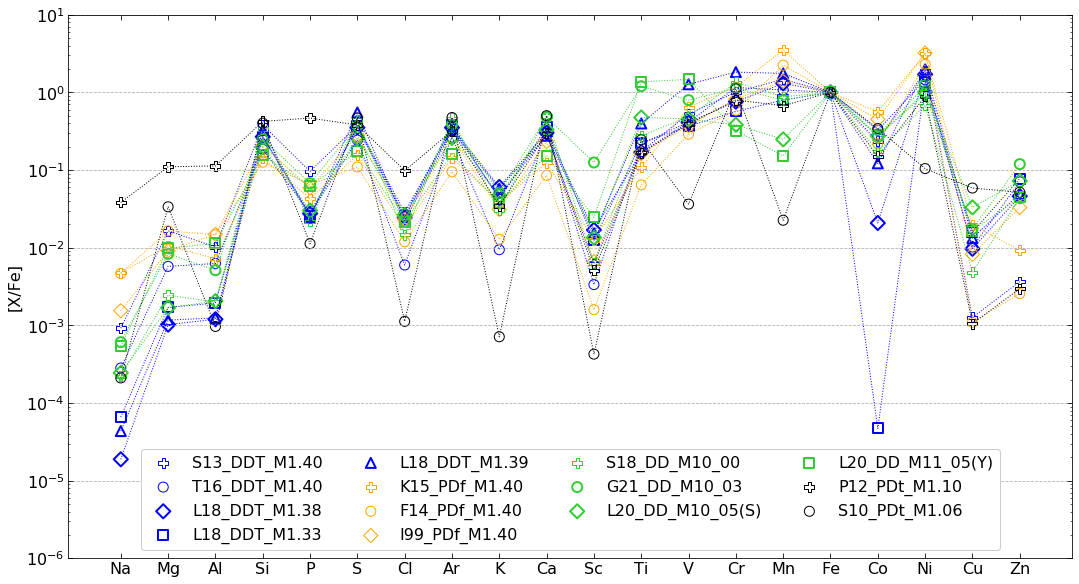} 
    \caption{Elemental yields normalised to Fe with respect to solar ratios \citep{Asplund:2021} for the SN Ia models investigated in this work: DDT (blue), PDf (orange), DD (green), and PDt (black). The models shown are based on an explosion of a WD progenitor with $Z=Z_{\odot}$.}
    \label{fig:yields}
\end{figure*}

Fig. \ref{fig:yields} shows the elemental yields relative to Fe for all models investigated in this work. Where yields have been calculated for several WD progenitors with different initial metallicities, only yields for the solar metallicity model are shown. The intermediate mass elements (IMEs) show a clear odd-even effect, with even-Z elements produced more efficiently \citep[e.g.][]{prantzos:2008, pagel:2009}. In Fig. \ref{fig:yields} there is an increase in the elemental yields of the Fe-group elements starting with Ti, which is often classified as an $\alpha$-element such as O or Si, but from a nucleosynthesis point of view should be classified as an Fe-group element \citep{Woosley:1995}.

\subsection{Deflagration-to-detonation transition (DDT) models}

In order for near-$M_{\text{Ch}}$ progenitors to produce the abundances of intermediate mass elements (e.g. Ne, Mg, Si, S, and Ca) that are observed in the spectra of SNe Ia, prior to detonation the fuel density must decrease, else the subsequent thermonuclear runaway will only produce isotopes at and around the Fe peak \citep{Arnett:1969, Steinmetz:1992}. The necessary expansion of the fuel occurs due to the energy released during a subsonic propagation of the combustion front at the onset of C-burning \citep{Ropke:2007}. After some time, once the fuel density falls below $\sim10^7\text{g cm}^{-3}$, the subsonic deflagration of the combustion front transitions into a supersonic detonation which produces enough nuclear energy to unbind the WD. These explosions are also often referred to as delayed-detonations, due to the period of pre-expansion of the fuel prior to detonation. 

In \cite{seitenzahl:12}, a three-dimensional hydrodynamic code is used to calculate yields for 14 models covering a parameter space of 11 different ignition scenarios. The ignition geometry of the deflagration is determined by the arrangement of the various numbers of spherical ignition kernels (ranging from 1 to 1600), where a higher number of kernels results in a higher degree of spherical symmetry. The ignition setup determines the strength of deflagration and the extent to which fuel can expand prior to the detonation. In this work, we use the model with 100 ignition kernels with a central density of $\rho_{c}=2.9\times10^9\text{g cm}^{-3}$, henceforth \citetalias{seitenzahl:12}. This model is post-processed by \cite{seitenzahl:12} at four different metallicities by reducing the $^{22}$Ne mass fraction to 0.01, 0.1, and 0.5 of the solar value ($M_{\text{Ne},\odot}=0.025$).

Using a two-dimensional hydrodynamic code, \cite{leung2018} calculate chemical yields resulting from DDT explosions in near-$M_{\text{Ch}}$ WDs. However, the authors note that the WD mass is extended down to $\sim1.30 M_{\odot}$ which could more closely resemble a sub-$M_{\text{Ch}}$ explosion. SN Ia yields are calculated by post-processing the explosions of WD progenitor covering a large range of model parameters, such as central density, metallicity, and flame shape. Herein, we use the yields from the benchmark model with $\rho_{c}=3\times10^{9}\text{g cm}^{-3}$ and $M_{WD}=1.38M_{\odot}$ at 7 different metallicities ($Z=0,0.1,0.5,1,2,3,5\: Z_{\odot}$), hereafter \citetalias{leung2018}8. Yields are also calculated at these metallicities for models with initial masses of $1.33M_{\odot}$ and $1.39M_{\odot}$, corresponding to central densities of $\rho_{c}=1\times10^{9}\text{g cm}^{-3}$ and $\rho_{c}=5\times10^{9}\text{g cm}^{-3}$ respectively. The \citetalias{leung2018}3 and \citetalias{leung2018}9 models present an excellent opportunity to explore the impact that central density/initial mass of the progenitor has on the GCE of the Fe-group elements. 

Finally, we include the DDT model of  \cite{townsley:16} with updated yields from \cite{Keegans:2023}. The yields are calculated by post-processing the trajectories of the \cite{townsley:16} model with an extended reaction network of over 5000 isotopes and 75,000 reactions. The model, hereafter \citetalias{townsley:16}, uses an artificially thickened reaction front in order for a cell in the computational grid to consist of both unburned and burned material. A reconstruction of the temperature-density history of these cells is then performed to approximate the passing of a fluid element through a reaction front of realistic size \citep{townsley:16}. The detonation is triggered by inserting a hot spot into the WD progenitor, with a central density at ignition of $\rho_{c}=2\times10^{9}\text{g cm}^{-3}$.

\subsection{Pure deflagration (PDf) models}

In contrast to the DDT scenario, if a detonation is not triggered and the combustion front only proceeds as deflagration then the nuclear energy released is insufficient to fully unbind the WD, leaving behind a bound remnant. Since in this regime the flame front propagates outwards at subsonic speeds, material further from the centre of the WD will have time to expand before being burnt and will therefore be synthesised mostly into intermediate mass elements \citep{Ropke:2007b}. Conversely, fuel near the centre will be burnt at higher densities producing mostly Fe-peak nuclei. The speed of the flame front determines the relative ratio of intermediate-mass to Fe-peak elements, with faster flame speeds producing relatively more of the last. In recent years, pure deflagration explosions have been proposed as potential candidates to explain the subluminous Type Iax (SN Iax) supernovae \citep{Jha:2017}. This class of SNe are thought to have short delay times since they are not readily observed in elliptical type galaxies \citep{Foley:2009, Valenti:2009}.

The most widely used SN Ia nucleosynthesis yields in GCE models of the Milky Way are those of \cite{iwamoto:99}, hereafter \citetalias{iwamoto:99}, based on the W7 pure deflagration model of \cite{Nomoto:1984}. The progenitor WD is constructed from a mass grid of 200 zones within an implicit Lagrangian hydrodynamics code, using reaction network data from \cite{Thielemann:1996}. The explosion is modelled starting from a WD of mass $M=1M_{\odot}$, which grows by accreting H-rich material at a steady rate of $\dot{M}_H=4\times10^{-8}M_{\odot}\text{y}^{-1}$ until carbon is ignited at a central density of $\rho_c=2.5\times10^9$ g cm$^{-3}$. Following this, the convective core undergoes thermonuclear runaway at $T_c \approx8\times10^8$ K and $\rho_c=2.12\times10^9$ g cm$^{-3}$, with a laminar flame propagation speed of $10\%-30\%$ of the sound speed, which results in a significant overproduction of $^{58}$Ni due to the fast burning front producing too much material in the range $Y_e=0.47-0.485$. This can be partially solved by reducing the metallicity of the zero-age main sequence progenitor from [Fe/H]$\;=0$ to [Fe/H]$\;=-1$ \citep[see Figure 6 in][]{Thielemann:2003}. Recently, the W7 model was post-processed by \cite{leung2018} with updated nuclear reaction network and electron capture rates; these revised yields provide a much better match with [Ni/Fe] at solar metallicity \citep{palla:2021}.

As in \cite{seitenzahl:12}, \cite{fink:2014} perform three-dimensional hydrodynamical simulations of an exploding WD for different numbers of ignition kernels. Although they use the same initial conditions as the former, the combustion front never becomes supersonic. As is the case for the DDT simulations in \cite{seitenzahl:12}, we use the yields from the model with 100 ignition kernels, hereafter \citetalias{fink:2014}, in our GCE analysis. 

\cite{kromer:15} simulate the explosion of a $1.4M_{\odot}$ hybrid WD with $\rho_{c}=2.9\times10^{9}\text{g cm}^{-3}$ following an off-centre deflagration, hereafter \citetalias{kromer:15}. The WD has a so-called ``hybrid" structure, comprising a CO core surrounded by an ONe mantle, and is formed as a late evolutionary stage of super asymptotic giant branch stars \citep{Denissenkov:2013}. The abrupt change in the composition profile of these stars is thought to lead to lower $^{56}$Ni yields, and as such they are postulated to be a potential progenitor scenario for the faintest SN Iax, such as SN 2008ha.      

\subsection{Double-detonation (DD) models}

The DD progenitor system involves the accretion of He-rich material onto the surface of a sub-$M_{\text{Ch}}$ WD from its binary companion via Roche-lobe overflow \citep{woosley:1994, Livne:1995}. The companion can be either a helium-burning star, a helium WD, or hybrid WD, with the single-degenerate channel giving rise to short delay times \citep{Ruiter:2011}. Regardless of whether the binary system is single- or double-degenerate, the detonation of CO material in the core of the WD is triggered by a preliminary He-shell detonation due to the compressional heating of the accreted material onto its surface.

In \cite{gronow:21b}, yields are calculated resulting from the DD explosions of WD progenitor with core masses from $0.8M_{\odot}$ to $1.1M_{\odot}$ and shell mass between $0.02M_{\odot}$ and $0.1M_{\odot}$. The temperature and density of the hydrodynamic simulations are mapped out by $2$ million tracer particles with a reaction network of $\sim30$ key isotopes. The explosion is then post-processed with an extended nuclear network to obtain yields for a primary WD progenitor with initial metallicity equal to solar metallicity. The work of \cite{gronow:21gce} extends the yield set to explosions from progenitor of initial metallicities of $0.01$, $0.1$ and $3Z_{\odot}$. In this work, we consider the yields from the M10\_03 ($1M_{\odot}$ core with $0.03M_{\odot}$ shell mass) model at all metallicities, hereafter \citetalias{gronow:21gce}, since the authors claim that this model best matches the solar values of Ti, V, and Cr.  

The DD explosions of sub-$M_{\text{Ch}}$ WD simulated in \cite{leung:20} are based on the same two dimensional hydrodynamic code used in \cite{leung2018}. Again, yields are calculated for a large parameter space of initial conditions including metallicity, mass, and the geometry of the He detonation (bubble, spherical, ring). Metallicity-dependent yields are calculated for three ``benchmark" models (one for each detonation trigger) that have the smallest He shell, whilst also being able to eject $\sim0.6M_{\odot}$$^{56}$Ni in line with most `normal' SN Ia observations. From these, we include the benchmark models with a ring (Y) and spherical (S) He detonation configuration. In a Y-type detonation, the He detonation is initiated as a ring around the rotation axis of the WD, whereas in an S-type detonation the He detonation is spherical and takes place 50 Km above the He/CO interface. Henceforth, we refer to these models as \citetalias{leung:20}\_M11\_05(Y) and \citetalias{leung:20}\_M10\_05(S), respectively. We choose to include these two benchmark models because they have an identical He envelope mass $M_{\text{He}}=0.05 M_{\odot}$, as opposed to $M_{\text{He}}=0.1 M_{\odot}$ for the bubble He detonation, therefore they offer a more direct probe of the effect of He detonation pattern, rather than $M_{\text{He}}$, on the GCE of the Fe-group elements. The effect of $M_{\text{He}}$ on the GCE can be seen by comparing the models of \cite{leung:20} to, for example, \citetalias{gronow:21gce} which has $M_{\text{He}}=0.03$ instead.   

The final set of DD yields investigated in this work are based on post-processing of the $1.0M_{\odot}$ model in \cite{shen:2018}, hereafter \citetalias{shen:2018}. The classification of this model as a DD detonation is somewhat tenuous, since the explosion is based on a ``bare" C/O WD with no He shell - the explosion is representative of a double-detonation of a CO WD resulting from the rapid accretion of material from a low-mass He WD companion \citep[see also][]{shen:2014}. The detonation in the core is ignited artificially by inserting a hot spot of $\sim2\times10^9$ at its core. For the purposes of this work we consider this model representative of a dynamically driven double detonation, where the He detonation triggers the C detonation but that the ashes of the latter dominate the yields since the He shell mass is negligible. We include in our GCE uptated yields from \cite{Keegans:2023}, calculated by post-processing the \citetalias{shen:2018} model and using the same nuclear reaction network as for \citetalias{townsley:16}.    

\subsection{Pure detonation (PDt) models}

Sub-$M_{\text{Ch}}$ WDs are proposed to explode as PDt if there is no helium-rich outer shell. A problem with most DD explosion models is the tendency to overproduce $^{56}$Ni due to the He-shell detonation prior to core detonation. However, whilst the absence of such a He-shell in theory produces light curves and spectra that can better match observations, it presents a new problem as to how the core detonation is triggered. One such solution is the violent merger of two WDs due to loss of angular momentum, which can nullify the need for a preliminary surface detonation of He in order to trigger a carbon detonation in the core. In this case, the material accreted onto the primary is rich in C+O rather than He which could lead to brief periods of C ignition on the WD surface. This class of SN Ia would appear less luminous than those sub-$M_{\text{Ch}}$ progenitor that experienced a He detonation, since in the latter large amounts of $^{56}$Ni are produced during burning in the He shell.

\cite{pakmor:12} model the explosion resulting from the violent merger of two WDs with masses 0.9 and 1.1 $M_{\odot}$, hereafter \citetalias{pakmor:12}. The binary system is evolved beginning with the coalescence of the stars and the subsequent detonation is triggered due to the heating and compression of material at the surface of the primary, which causes hot spots to form deeper inside the star. The nucleosynthesis is calculated by post processing the temperature and density profile recorded by $10^6$ tracer particles.

Although the violent merger scenario is a plausible site for PDt progenitor, the resulting mass of the WD remnant could exceed $M_{\text{Ch}}$ in the case where the mass ratio of the merging WDs is $\lesssim0.8$ \citep{pakmor:2011, sato:2016} leading to an accretion-induced collapse of the WD remnant to a neutron star \citep[see e.g.][]{liu:2020}. The exact physical conditions that determine whether a merging binary system of two CO WDs results in an accretion-induced collapse or violent merger are still not fully clear. Therefore, it is not uncommon for models of PDt SN Ia explosions to remove the ambiguity surrounding the detonation trigger and instead concern themselves only with the timeframe starting from core detonation. This latter approach is adopted by \cite{sim:2010}, who use one-dimensional hydrostatic models to explore PD explosions in WDs with masses from 0.97 to 1.15 $M_{\odot}$. The models consider only the explosion in a numerical context without consideration of the underlying trigger for the detonation. In this work, we use the yields from the 1.06 $M_{\odot}$ C+O WD model, hereafter \citetalias{sim:2010}, since its light curve is in good agreement with observations. We choose not to use the 1.06 $M_{\odot}$ C+O+Ne WD with higher initial Ne composition as this model produces a higher abundance of Fe-group elements but comparatively less intermediate mass elements. 

\section{The chemical evolution code}\label{sec:GCE}

We use the \texttt{OMEGA+} GCE code \citep{cote17, cote18} to model the evolution of chemical abundances in the Galaxy. The GCE framework is calibrated based on the Milky Way's disk and takes into account present day observational estimates for the star formation rate $(\text{SFR; }M_{\star})$, the gas infall rate, supernovae rates, and the total mass of gas $(M_{\text{gas}})$ in the Galaxy. \texttt{OMEGA+} is a two-zone model comprised of a central cold gas reservoir with active star formation that contains all stellar populations in the Galaxy; this is encompassed by a hot gas reservoir devoid of star formation and stellar populations. These zones are referred to henceforth as the ``galaxy" and the ``circumgalactic medium" respectively, and gas can circulate between the two zones via galactic inflows and outflows. The change in the total mass of gas in the Galaxy at time $t$ is

\begin{equation}
    \dot{M}_{\text{gas}}(t) = \dot{M}_{\text{inflow}}(t) + \dot{M}_{\text{ej}}(t) - \dot{M}_{\star}(t) - \dot{M}_{\text{outflow}}(t),
\end{equation}
where the terms on the right-hand side of the equation are, from left to right, the gas inflow rate into the Galaxy from the circumgalactic medium, the rate at which gas is returned to the ISM from stellar ejecta, the star formation rate (SFR), and the rate of gas outflow from the Galaxy into the circumgalactic medium due to stellar feedback. Below we describe each of these four terms in more detail. 

Beginning firstly with the gas inflow rate $(\dot{M}_{\text{inflow}})$, for all of our GCE calculations in this work we assume that the formation of the Milk Way disk can be attributed to two distinct infall episodes. In this two infall model, the gas inflow rate $(\dot{M}_{\text{inflow}})$ is given by \citep[][]{chiappini:1997} 

\begin{equation}
    \dot{M}_{\text {inflow }}(t)=A_{1} \exp \left(\frac{-t}{\tau_{1}}\right)+A_{2} \exp \left(\frac{t_{\max }-t}{\tau_{2}}\right),
\end{equation}
where each exponential term describes a separate infall episode. The second infall is delayed by $t_{\text{max}}=1.0$ Gyr with respect to the first and occurs over a much longer timescale: $\tau_{1}\simeq0.7$ Gyr for the first infall compared to $\tau_{2}\simeq7$ Gyr for the second. The normalisation constants $A_1$ and $A_2$ are chosen so as to best reproduce the current estimated mass of gas in the Galactic disk as derived by \cite{kubryk:2015}. 

Mechanical energy released from CCSNe can propel gas out of the Galaxy into the CGM. Since these galactic outlflows are driven by stellar feedback, the outflow rate $\dot{M}_{\text{outflow}}$ is proportional to the SFR, such that 

\begin{equation}
    \dot{M}_{\text{outflow}}(t)=\eta \dot{M}_{\star}(t).
\end{equation}
The SFR ($\dot{M}_{\star}$) is a linear function of $M_{\text{gas}}$,

\begin{equation}
    \dot{M}_{\star}(t) = f_{\star}M_{\text{gas}}(t),
\end{equation}
where $f_{\star}$ represents the star formation efficiency in units of $[\text{yr}^{-1}]$. At each timestep, the \texttt{OMEGA} code converts some amount of the gas in the Galaxy into a simple stellar population (SSP). The mass of the SSP is determined by the current star formation rate and the distribution of stellar masses it contains is based on the chosen initial mass function (IMF); in this work we adopt the IMF of \cite{kroupa:1993}. All stars in a given SSP are assumed to be born simultaneously with a chemical composition identical to that of the gas reservoir at that time. However, the stars in an SSP will eject their yields into the ISM over different timescales depending on the specific delay-time distribution function. The Stellar Yields for Galactic Modeling Applications \citep[SYGMA;][]{rit18} module is called by \texttt{OMEGA} at each timestep to calculate the composition of the stellar ejecta from all SSPs in the system based on the mass- and metallicity-dependent stellar yields being used. 

In this work we include in our GCE framework the yields of low- and intermediate-mass stars (LIMS) and massive stars, together with the SN Ia yields described in Section \ref{sec:yields}. The LIMS yields of \cite{cristallo:2015} are used in all of our GCE calculations, however, we perform two sets of independent GCE calculations using the massive star yields of \cite{limongi2018} (hereafter LC18) and \cite{nomoto13} (hereafter N13). For LC18, the yields are weighted according to a metallicity-dependent function of the rotational velocity as derived by \cite{prantzos:18}, and we include in the IMF massive stars up to $100M_{\odot}$, however, stars more massive than $25M_{\odot}$ are assumed to collapse to a black hole and thus contribute to chemical enrichment only by virtue of their winds \citep[i.e. set R in][]{limongi2018}. The explosion of the star is initiated by means of a ``kinetic bomb", where at some interior mass coordinate an expansion velocity is imparted that is sufficient to entirely eject material exterior to the Fe core. For the N13 yields, we do not assume any contribution from hypernovae, and the upper limit for the IMF is restricted to $40M_{\odot}$ so that $Z=Z_{\odot}=0.014$ at [Fe/H]$\;=0$ \citep[see also][where the upper mass limit is also curtailed]{gronow:21gce}. In contrast to LC18, the explosions are triggered using a ``thermal bomb" -- that is, the outwardly propagating shockwave is driven by artificially inserting a hot spot into the innermost regions of the star. In all GCE models the transition mass between LIMS and massive stars is set to $8M_{\odot}$ \citep[see, e.g.][]{karakas:14}. We choose to run GCE calculations for both sets of massive star yields because in \cite{gronow:21gce} it was found that the choice of massive star yields has large implications for the relative contributions from sub- and near-$M_{\text{Ch}}$ SN Ia progenitors towards the solar Mn abundance.

We adopt a fiducial delay-time distribution (DTD) for SNe Ia of the form $t^{-1.07}$ as derived by \cite{maoz:12}. This DTD is normalised so that $\sim 10^{-3}$ SNe form per stellar mass in a SSP, which is to the same order of magnitude as rates derived from observations \citep[for comparison, a compilation of observed Galactic SN Ia rates are given in Table 5 of][]{cote:2016}. A power law form for the DTD is expected for double degenerate systems in which the delay time is dominated by the merger timescale, assuming the distribution for the orbital separation $S$ of binary stars follows $f(S)dS\sim S^{-1}$ \citep[see][]{ruiter:09}. Indeed, a power law DTD with exponent $\approx -1$ has also been found to have the best fit to the observed SN Ia rate in nature \citep{Totani:2008, maoz:12}. On the other hand, delay times for single degenerate systems will depend mostly on the evolutionary timescale of the donor \citep{ruiter:09}. However, \cite{matteucci:2009} demonstrated that DTDs calculated using either single or double degenerate scenarios lead to a very similar chemical evolution of [O/Fe] vs [Fe/H]. Similarly, we find negligible difference for the GCE of Fe-group elements when using instead the DTD calculated by \cite{ruiter:14} for a WD accreting from a non-degenerate companion.

\section{Galactic chemical evolution predictions}\label{sec:results}

\subsection{Comparison to observations for individual SN Ia yields}\label{sec:GCE_vs_obs}

\begin{figure*}[!h]
    \centering
   
    \includegraphics[width=1.0\textwidth]{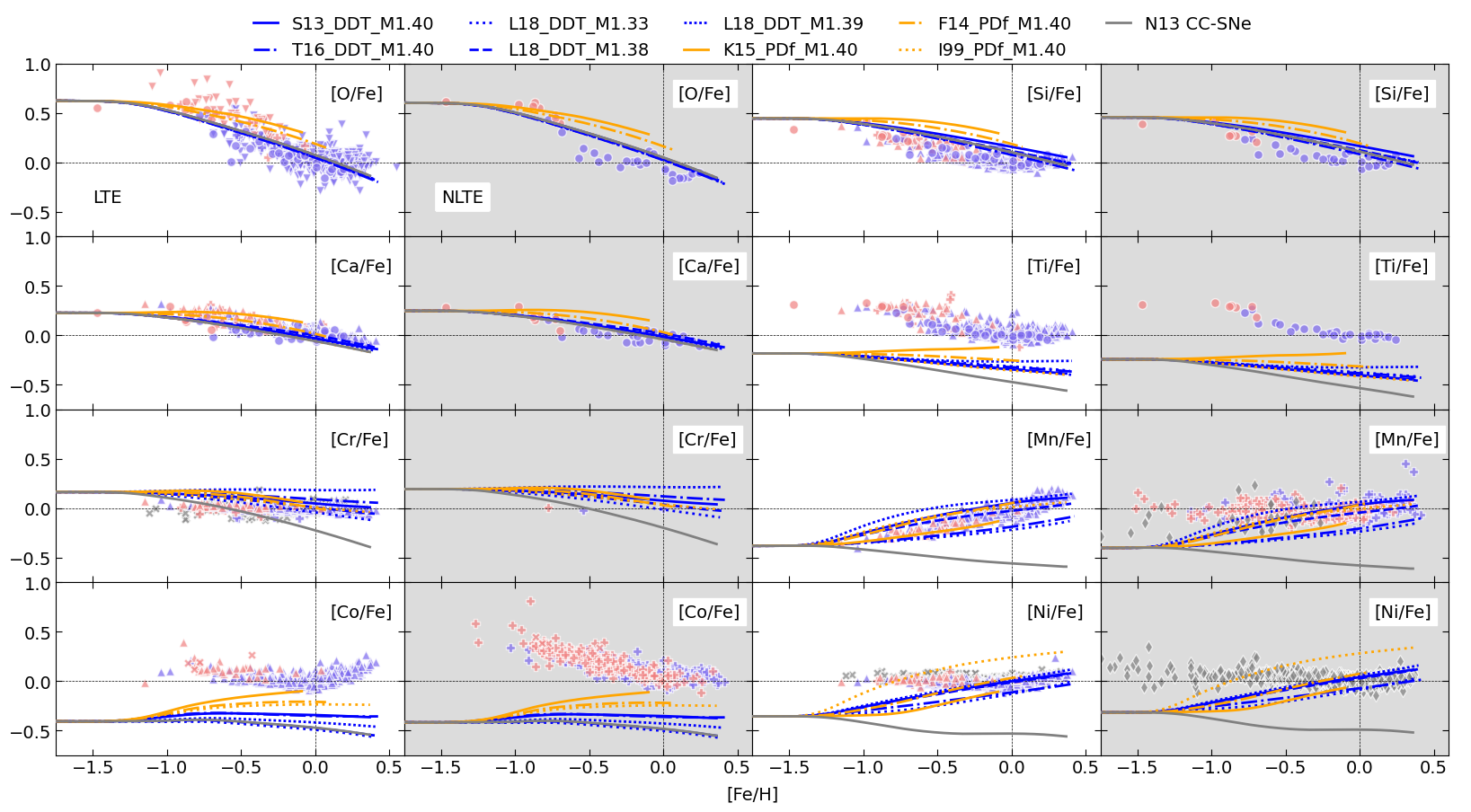} 
    \includegraphics[width=1.0\textwidth]{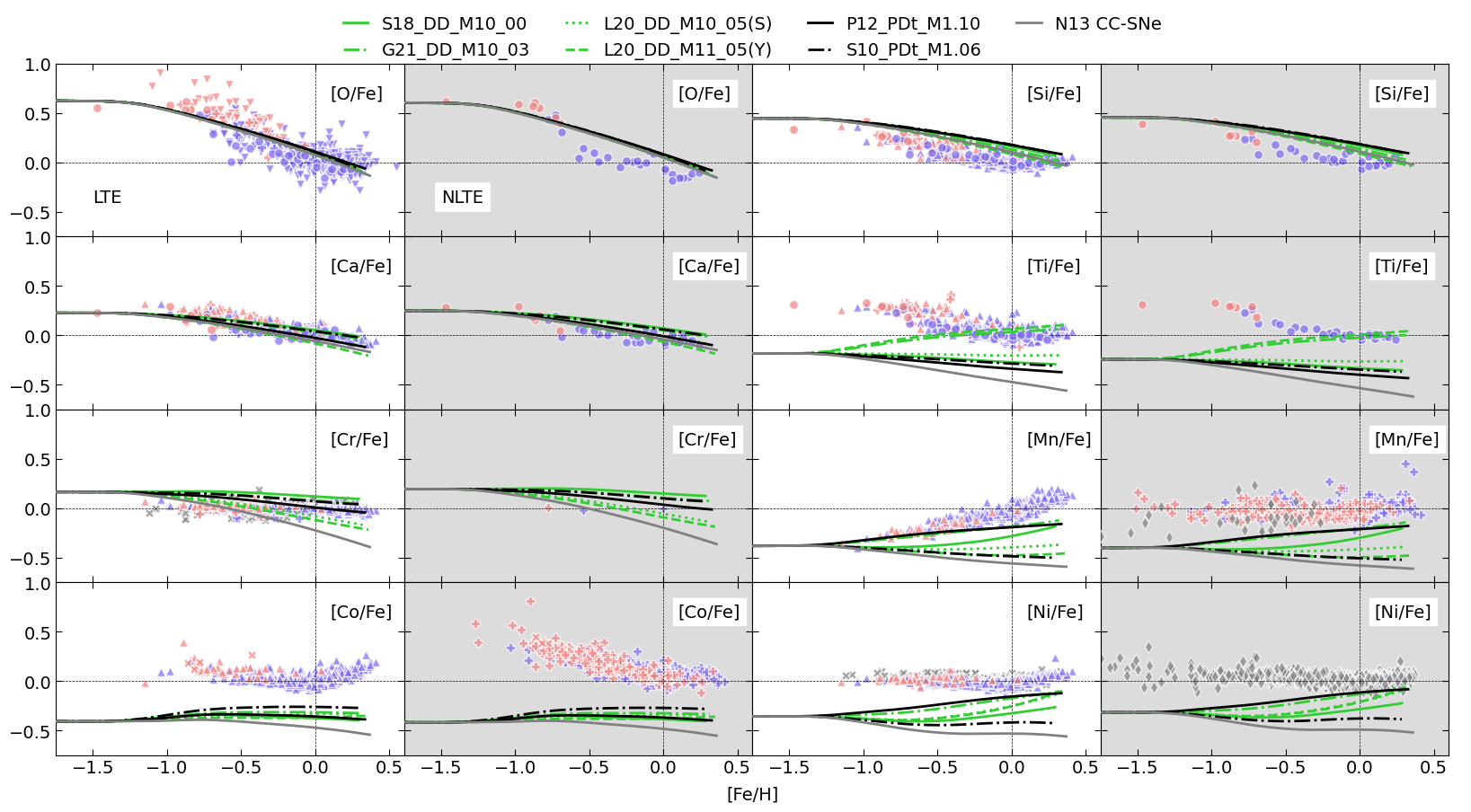} 
    \caption{Predicted evolution of [X/Fe] versus [Fe/H] in the Milky Way's disk for O, Ca, Si, and the Fe-peak elements for the GCE models with N13 CCSN yields. The top and bottom panels show predictions with near- and sub-$M_{\text{Ch}}$ SN Ia yields, respectively. For each element, LTE (panels without shading) and NLTE (shaded) observational data are plotted separately. The GCE tracks are colour coded according to the explosion mechanism (see text), where the grey line shows the evolution of [X/Fe] with no contribution from SNe Ia (i.e. assuming only a contribution to X from CCSNe). See Table \ref{table:data} for the list of observational datasets used.}
    
    \label{fig:N13_GCE}
\end{figure*}

\begin{figure*}[!h]
    \centering
    \includegraphics[width=1.0\textwidth]{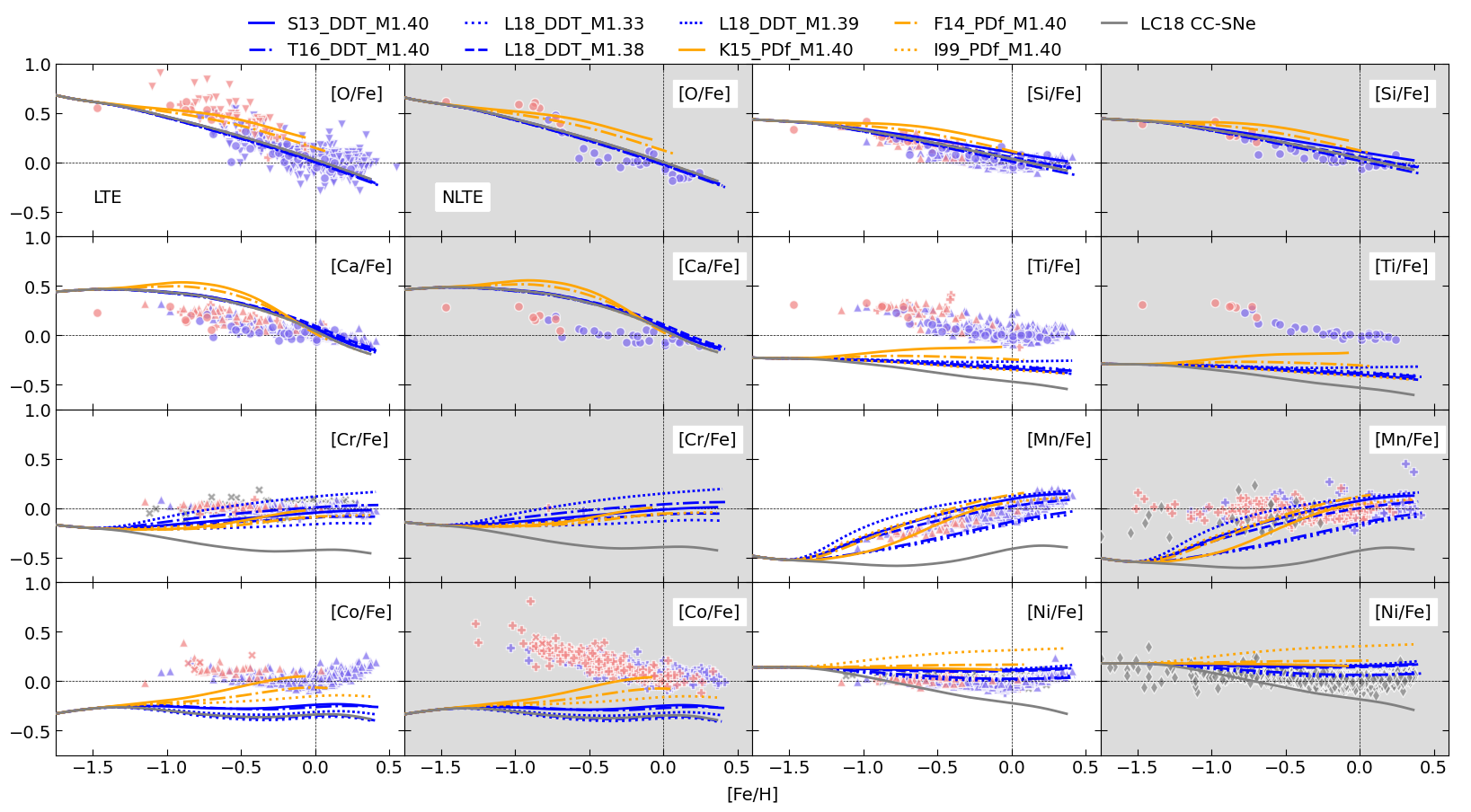} 
    \includegraphics[width=1.0\textwidth]{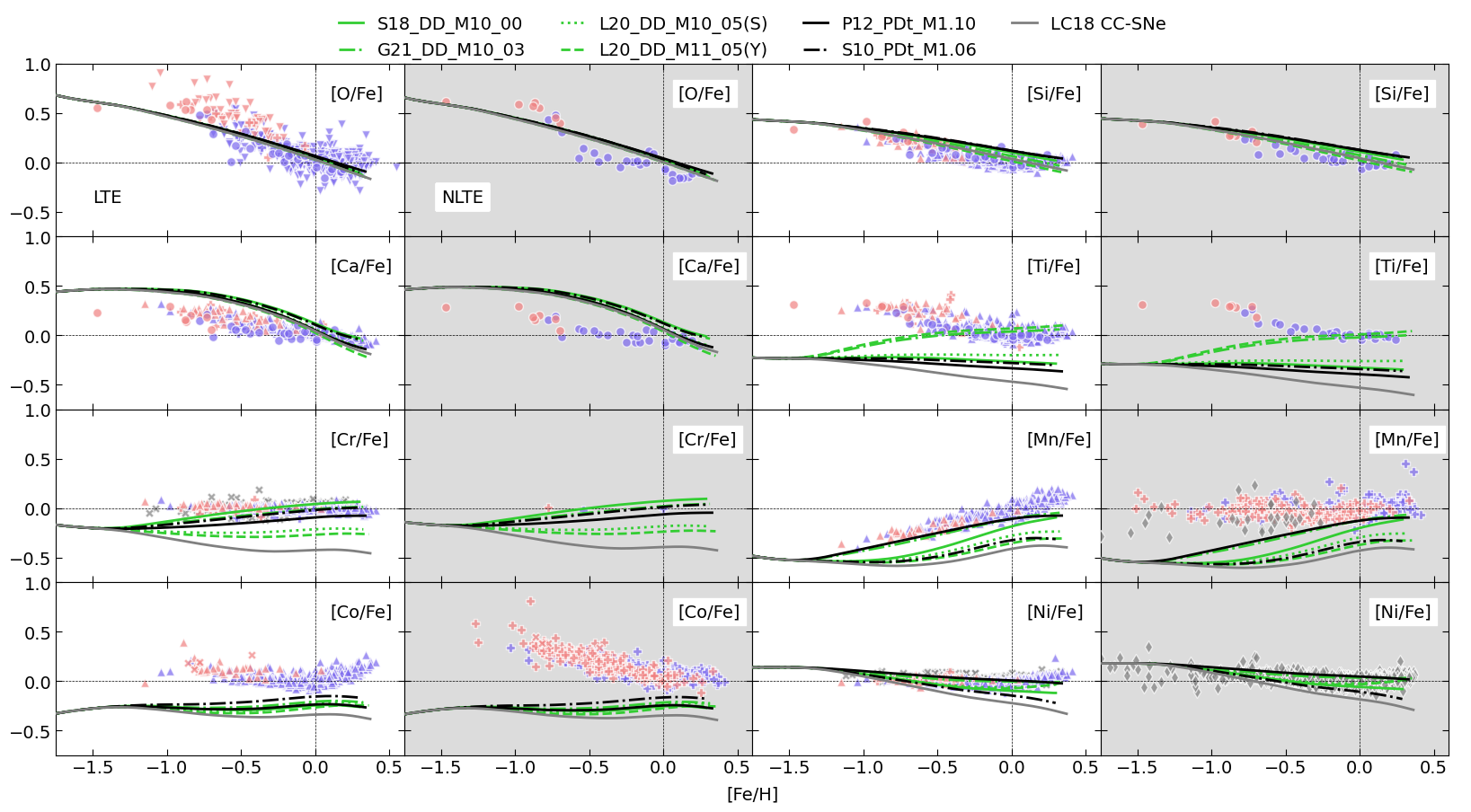}
    \caption{Same as Fig. \ref{fig:N13_GCE} but for LC18 CCSN yields.}
    \label{fig:LC18_GCE}
\end{figure*}

Fig. \ref{fig:N13_GCE} shows the evolution of [O, Si, Ca, Ti, Cr, Mn, Co, Ni/Fe] versus [Fe/H] for the Milky Way disk as predicted by our GCE model with N13 massive star yields. Fig. \ref{fig:LC18_GCE} shows the same but for LC18 massive star yields. Panels shaded in grey contain only NLTE data, whereas those that are unshaded contain only LTE data. The GCE predictions are normalised to the solar values of \cite{Asplund:2021} and \cite{Grevesse:2007} for NLTE and LTE data, respectively. The top (bottom) set of panels includes SN Ia yields from near (sub)-$M_{\text{Ch}}$ WD progenitor and the lines are colour coded according to the explosion mechanism: DDT (blue), PDf (orange), DD (green), and PDt (black). The GCE code is calibrated to reach [Fe/H]$\;\approx0$ at the time of the birth of the Sun ($t_{\odot}=8.6$ Gyr) based on global properties of the Milk Way's disk (e.g. star formation rate, gas inflow rate, mass of gas, etc.) and current estimates for supernovae rates (both SNe Ia and CCSNe). However, due to the fact a bound remnant remains following the PDf explosions of \citetalias{fink:2014} and \citetalias{kromer:15}, less material is ejected from these SNe Ia and so the associated GCE tracks have a lower final [Fe/H] than the other models. In addition to the GCE predictions that include an SN Ia contribution, we show with a grey line the evolution of [X/Fe] but with no contribution of X from SNe Ia (i.e. only the contribution from massive stars). For those datasets that are delineated into different stellar populations, or datasets where recommended parameters for such are given, thick and thin disk stars are plotted separately in red and blue, respectively. Otherwise, disk stars are plotted in grey.

\setlength{\arrayrulewidth}{0.3mm}
\renewcommand{\arraystretch}{1.5}

\begin{table*}
\caption{List of LTE and NLTE observational datasets used for each element.}
\centering
\begin{tabular}{c|P{5cm}|P{5cm}}
    Element & LTE & NLTE \\
    \hline
    O & \cite{Bensby:2005}, \cite{Bertran:2015}, \cite{Zhao:2016} & \cite{Zhao:2016} \\
    Si, Ca, Ti & \cite{Bensby:2005}, \cite{adibekyan:2012}, \cite{Zhao:2016} & \cite{Zhao:2016} \\
    Cr & \cite{Bensby:2005}, \cite{adibekyan:2012}, \cite{Bergemann:2010}, \cite{Lomaeva:2019} & \cite{Bergemann:2010} \\
    Mn & \cite{adibekyan:2012} & \cite{Battistini:2015}, \cite{eitner:2020} \\
    Co & \cite{Bergemann:2010}, \cite{adibekyan:2012} & \cite{Bergemann:2010}, \cite{Battistini:2015}, \\
    Ni & \cite{Bensby:2005}, \cite{adibekyan:2012}, \cite{Lomaeva:2019} & \cite{Eitner:23}\\
\end{tabular}
\label{table:data}
\end{table*}

\subsubsection{The $\alpha$-elements (O, Si, Ca)}

While O is only produced by massive stars, our GCE models indicate that SNe Ia contribute around $15-20\%$ of the Solar Si and Ca. The NLTE corrections follow the same general trend for the [O, Si, Ca/Fe] data -- at [Fe/H]$\;<-1.0$ the corrections are mostly positive but are negative at higher metallicities. For Si and Ca the corrections are small, generally confined to $<|0.1|$ dex at the metallicity limits plotted in Fig. \ref{fig:N13_GCE} \& \ref{fig:LC18_GCE}. The NLTE corrections are larger for O, reaching NLTE $-$ LTE $\gtrsim|0.2|$ dex for [Fe/H]$>0.5$. The metallicity trend of the $\alpha$-elements is well know, having a well defined super-solar plateau in the low-metallicity regime due to the ejecta of massive stars, which then turns downwards sharply after [Fe/H]$\;>-1$ as SNe Ia begin to contribute large amounts of Fe \citep[e.g.][]{matteucci:1986}. 

The [O/Fe] evolution is generally well reproduced by the GCE models. In particular, the N13 models best fit the thick disk but slightly overproduce [O/Fe] relative to the thin disk, whereas the LC18 models slightly underproduce at low metallicities but better fit the higher metallicity data. The PDf models do not exhibit as sharp of a ``knee'' at [Fe/H]$=-1.0$ due to the lower amounts of Fe ejected compared to the other explosion types. The N13 models slightly overproduce [Si/Fe] compared to observations, however, the LC18 models fit the LTE and NLTE data well. The [Ca/Fe] is $\sim0.15$ dex higher from LC18 CCSNe than N13, which leads to the latter better reproducing the low-metallicity plateau and bulk of the thick disk observations at [Ca/Fe]$\;\approx0.35$. Due to the burning at comparatively lower central densities in the PDf models, the \citetalias{kromer:15} and \citetalias{fink:2014} GCE tracks with LC18 CCSNe initially show a small increase from the plateau value of [$\alpha$/Fe] after [Fe/H]$\;=-1$. Furthermore, since these two models eject comparatively lower amounts of Fe, they also have higher final [$\alpha$/Fe] abundance ratios than other explosion types. The other GCE tracks show little variation between explosion types and masses of the WD progenitors.

\subsubsection{Titanium}

The observed Ti versus metallicity relationship is well known to mimic that of the lighter $\alpha$-elements. However, it can also be classified as an Fe-peak element from a nucleosynthesis standpoint since it is synthesised in the same region as Fe during a SN \citep{Keegans:2023}. As first determined by \cite{Bergemann:2011}, the NLTE corrections for Ti I can be in excess of 0.1 dex and therefore it is safer to use Ti II lines in GCE analyses. The NLTE correction to Ti II lines are negative but very small, typically less than 0.02 dex \citep{Zhao:2016}, and so the following analysis for the comparisons to GCE are applicable to both LTE and NLTE datasets. Whilst this paper was in its final stages \cite{Mallinson:2024} was published, which provides 1D NLTE effects on Ti abundances for a large sample of stars. They find that NLTE corrections are generally large and positive at very low-metallicities, which further exacerbates the current dichotomy between observations and GCE predictions.

Regardless of whether massive star yields are taken from N13 or LC18, the GCE predictions underproduce [Ti/Fe] relative to observations, at least prior to SN Ia enrichment in the ISM; this is a known phenomena of GCE models \citep[see, e.g.][]{kobayashi:2006, Hughes:2008, Romano:2010, Mishenina:2017, Kobayashi:2020a}. This is due to CCSN models currently underproducing $^{48}$Ti, that is the most abundant Ti isotope \citep{timmes:1995}. In particular, the production of its parent radioactive nucleus, $^{48}$Cr, is sensitive to the SN explosion conditions. For instance, \cite{Leung:2023} highlighted that explosions with a strong $\alpha$-rich freezeout, such as a jet-driven CCSNe, could yield high Ti enrichment and therefore be significant contributors of Ti in the Galaxy. The inclusion of such asymmetrical explosion geometries in GCE models could provide a better fit to the [Ti/Fe] data \citep{Kobayashi:2020a}, but these events are still fairly poorly constrained.

In both Figures \ref{fig:N13_GCE} \& \ref{fig:LC18_GCE}, all DDT models (blue lines) initially exhibit a steady decrease in [Ti/Fe] with [Fe/H], with only the \citetalias{leung2018}9 model plateauing at [Fe/H]$\;\sim-0.3$. The PDF models (orange lines) all behave slightly differently: for \citetalias{kromer:15} [Ti/Fe] steadily increases, with the rise becoming more gradual over time; \citetalias{iwamoto:99} decreases almost linearly with metallicity; \citetalias{fink:2014} rises at first before decreasing. The PDt models (black lines) show a similar trend to the DDT models. There is a large spread in [Ti/Fe] for the DD models (green lines), with both \citetalias{gronow:21gce} and \citetalias{leung:20}\_M11\_05(Y) approaching solar [Ti/Fe], whereas \citetalias{shen:2018} and \citetalias{leung:20}\_M11\_05(S) remain below [Ti/Fe]$\;\sim-0.2$ across the whole metallicity range. At [Fe/H]$\;=0$ there is a $\sim0.3$ dex difference in [Ti/Fe] for \citetalias{leung:20}\_M11\_05(Y) and \citetalias{shen:2018}. The discrepancies between the DD models for Ti can be explained by considering the fact that the most abundant isotope of Ti in the ejecta of these models, irrespective of the metallicity of the progenitor, is $^{48}$Ti. At higher temperatures ($>8$ GK) $^{48}$Ti is directly produced, however, at lower temperatures the largest contribution is from the decay of $^{48}$Cr \citep{Keegans:2023}. Since this radiogenic contribution will depend strongly on the burning that takes place during He detonation, the production of Ti will be higher for models with larger $M_{\text{He}}$. This explains why \citetalias{leung:20}\_M11\_05(Y) can produce supersolar amounts of Ti, and why the bare (i.e. no He shell) CO WD model \citetalias{shen:2018} behaves so similar to the PDt models for this element.

\subsubsection{Chromium}

The LTE data for [Cr/Fe] is mostly flat and close to the solar value at all metallicities. The NLTE data from \cite{Bergemann:2010} is only calculated for two disk stars and the Sun, and at present closely resembles the flat metallicity trend derived using LTE analysis. However, the discrepancy between LTE Cr I and NLTE Cr I data increases at lower metallicities, up to about 0.2 dex at [Fe/H]$\;-1.2$ \citep[][]{Bergemann:2010} within the metallicity range considered in this work; this could lead to a non-flat pattern for NLTE Cr once more data becomes available. Therefore, caution should be maintained for the GCE results fitting Cr, and a new analysis should be done once the NLTE [Cr/Fe] pattern with respect to metallicity is well constrained by observations.

Comparing Figures \ref{fig:N13_GCE} and \ref{fig:LC18_GCE} we see disparities between Cr production for the GCE calculations made using the two sets of massive star yields - there is a $\sim0.4$ dex higher [Cr/Fe] abundance at [Fe/H]$\;<-1.5$ for the GCE predictions using N13 massive star yields. Despite this, the majority of GCE tracks, irrespective of the choice of massive star yields, are in good agreement with the observational data after [Fe/H]$\;\gtrsim-1$ (notably, the yield combination of LC18 and \citetalias{leung:20} models remain subsolar at all metallicities).

In Fig. \ref{fig:N13_GCE}, the onset of SN Ia enrichment is met by a steepening drop in [Cr/Fe] with increasing [Fe/H]. Neither the detonation mechanism nor the mass of the WD progenitor seem to have a pronounced effect on the rate of decrease. In Fig. \ref{fig:LC18_GCE}, the predicted [Cr/Fe] increases and then begins to plateau around solar metallicity.

In contrast to the evolution of Ti and V, there is a negligible difference in the Cr evolution for models with different He detonation geometries (i.e. models \citetalias{leung:20}\_M11\_05(Y) and \citetalias{leung:20}\_M10\_05(S)). In fact these models underproduce Cr compared to the two other DD models, \citetalias{shen:2018} and \citetalias{gronow:21gce}, which initially suggests that a prominent production channel of Cr that is included in the reaction networks of \cite{Keegans:2023}\footnote{We remind that the yields for \citetalias{shen:2018} are updated in \cite{Keegans:2023} by post-processing the model with a more extensive network} and \cite{gronow:21b} is absent in the reaction network of \cite{leung:20}. However, \cite{palla:2021} find that Cr production is increased for the \cite{leung:20} benchmark model with aspherical detonation pattern, resulting in super-solar abundances of [Cr/Fe] - the S-type detonation produces sub-solar abundances as found in this work. In \cite{palla:2021} the aspherical yields are taken from the \cite{leung:20} benchmark model with X-type He detonation, which has a He shell twice as massive (i.e. $M_{\text{He}}=0.1M_{\odot}$) as those of \citetalias{leung:20}\_M11\_05(S) and \citetalias{leung:20}\_M11\_05(Y) used in this work. The significance of the He shell detonation on Cr production is called into question however, when we consider that the \citetalias{shen:2018} model with no He shell, as well as all PDt models, can produce [Cr/Fe] values that agree well with the observational data.

The complexity of Cr production can be somewhat explained by considering that although Cr has four stable isotopes ($^{50,52,53,43}$Cr), the solar abundance of this element comes almost entirely from the contribution of $^{52}$Cr ($\sim84\%$ of solar). The bulk of $^{52}$Cr production, regardless of progenitor mass, is as the radiogenic $^{52}$Fe at intermediate temperatures $\sim5-6$ GK during incomplete Si burning \citep{iwamoto:99, Keegans:2023}. This means that the Cr yield has very little dependency on the initial metallicity of the progenitor. Instead, Cr production is sensitive to conditions affecting nucleosynthesis during the explosion \citep{Keegans:2023}. This is particularly true when we compare [Cr/Fe] for \citetalias{leung2018}3, \citetalias{leung2018}8 and \citetalias{leung2018}9, which have different central densities. At higher metallicities, the high density model leads to an overproduction relative to the solar value, whereas [Cr/Fe] is underproduced by the low density model. Finally, we note that all PDf models produce [Cr/Fe] that is consistent with observations. 

\subsubsection{Manganese}

A key feature of Mn production in SNe Ia is the dependency on progenitor mass. For this reason, the [Mn/Fe] ratio has been used extensively as a tool to constrain the number of WD progenitor in the Galaxy of either a sub-$M_{\text{Ch}}$ or near-$M_{\text{Ch}}$ origin.

There are large differences in the observed evolution of [Mn/Fe] in disk stars depending on whether NLTE corrections are applied to the data or not. The data based on LTE analysis is highly sub-solar at low metallicities, and shows [Mn/Fe] only increasing after [Fe/H]$\;=-1$. However, recent studies have shown that Mn I lines in particular are strongly affected by NLTE effects \citep{bergemann:2019}. NLTE observations of disk stars suggest that [Mn/Fe] does not strongly evolve with metallicity, but instead remains close to the solar value even at [Fe/H]$\;<-1.5$. As can be seen in Figures \ref{fig:N13_GCE} and \ref{fig:LC18_GCE}, the GCE tracks more closely fit the LTE rather than NLTE data at low metallicities. Since CCSNe dominate the Mn production in the Milky Way at [Fe/H]$\;\lesssim-1$, a potential solution to fit the NLTE data at low metallicities is to assume an enhanced production of Mn from CCSNe \citep{eitner:2020, palla:2021}. However, in order to remain self-consistent, if the Mn yield is artificially increased then considerations must also be made regarding changes in the yields of other iron-group elements such as Ti and Cr, as well as $\alpha$-elements such as Ca that are produced in the same conditions and by the same nucleosynthesis processes as Mn in CCSN ejecta \citep{Woosley:1995, Thielemann:1996, Chieffi:1998}. For the GCE models in this work we prefer to consider only the unmodified yields as calculated by stellar nucleosynthesis models, but we acknowledge that for some elements this self-consistent approach will lead to less than desirable fits to the observational data.

The only stable isotope of Mn is $^{55}$Mn, which is predominantly made as the radioactive $^{55}$Co during deflagration and freeze-out from NSE, assuming small mass fractions of $^{4}$He \citep{seitenzahl:Mn}. At densities below $\rho\sim2\times10^8 \text{g cm}^{-3}$, $^{55}$Co is destroyed via the reaction $^{55}$Co(p,$\gamma$)$^{56}$Ni, which leads to a lower production of Mn from sub-$M_{\text{Ch}}$ progenitor. In Figures \ref{fig:N13_GCE} and \ref{fig:LC18_GCE} supersolar [Mn/Fe] at [Fe/H]$\;>0$ is obtained for GCE models with pure deflagration explosions, or SNe Ia with prolonged periods of burning with subsonic propagation of the flame front, such as the \citetalias{leung2018}9 model with higher central density. The ejected mass of Mn also increases with metallicity, since at higher metallicities $^{55}$Fe is more readily produced at temperatures of $3-4$ GK \citep{Keegans:2023}. This metallicity dependency is more pronounced for sub-$M_{\text{Ch}}$ progenitor than for near-$M_{\text{Ch}}$. In regards to this last, it is worth mentioning that in GCE models that consider only a contribution from near-$M_{\text{Ch}}$ it has been demonstrated that metallicity-dependent SN Ia yields are necessary to reproduce the [Mn/Fe] trends in dwarf spheroidal galaxies \citep{Cescutti:2008}.

The GCE models with LC18 CCSNe have a low-metallicity plateau of [Mn/Fe]$\;\approx-0.5$, which is $\sim0.1$ dex lower than those with N13. However, after [Fe/H]$\;\gtrsim-0.5$ extrapolation of the solar metallicity yields from the LC18 CCSN models causes a substantial increase in the Mn contribution compared to N13, resulting in a faster evolution to the solar [Mn/Fe] value for the majority of the near-$M_{\text{Ch}}$ GCE tracks. The disparity between CCSN yields is also noticeable when comparing the sub-$M_{\text{Ch}}$ GCE tracks between Figures \ref{fig:N13_GCE} \& \ref{fig:LC18_GCE}, since the high metallicity contribution from the LC18 CCSNe is significant enough to cause an upward trend in all of the tracks. There are also relevant differences between SN Ia models with the same explosion mechanism. For example, there is a $\sim0.3$ dex difference in [Mn/Fe] at [Fe/H]$\;=0$ for \citetalias{leung:20}\_M11\_05(Y) and \citetalias{gronow:21gce}, despite the fact these models are both based on a double-detonation explosion of a sub-$M_{\text{Ch}}$ WD. Likewise, there is a $\sim0.3$ dex difference between the delayed detonation models of \citetalias{townsley:16} and \citetalias{leung2018}9 at solar metallicity. These results confirm that GCE models cannot reliably use the [Mn/Fe] ratio to constrain WD progenitor mass in the Galaxy
without considering variations in the sub- and near-$M_{\text{Ch}}$ SN Ia yields.

Another important consideration is the extent of the contribution towards Galactic Mn from SN Iax. The deflagration models can produce supersolar [Mn/Fe], leading to [Mn/Fe]$\;\sim0.2$ in Fig. \ref{fig:LC18_GCE}. This subclass of SNe Ia could occur with a rate of up to $\sim30\%$ of the classical SN Ia rate \citep{srivastav:2022}, which could further reduce the necessary fraction of sub-$M_{\text{Ch}}$ progenitor. Population synthesis simulations of He-rich accretion in the single-degenerate scenario predict that the majority of SN Iax should explode in the order of $\sim10^{7}$ years \citep{liu:2015}. A prompt enrichment of supersolar [Mn/Fe] from SN Iax could help to account for the higher [Mn/Fe] data at low metallacities when accounting for NLTE corrections \citep{kobayashi:2015}.

\subsubsection{Cobalt}

The majority of Co in the Galaxy is made by CCSNe, with a contribution of $\sim20\%$ from SNe Ia at [Fe/H]$\;=0$. In both Figures \ref{fig:N13_GCE} \& \ref{fig:LC18_GCE} the GCE tracks have [Co/Fe]$\;=-0.2$ at [Fe/H]$\;=-1.5$, which indicates that there is little difference in Co production from N13 and LC18 CCSNe at lower metallicities. 

The NLTE data shows a decreasing trend with metallicity, whereas the LTE data are mostly flat between [Fe/H]$\;>-1$ and solar metallicity, beyond which [Co/Fe] increases steadily in the super-solar metallicity regime. The GCE predictions do not quantitatively or indeed even qualitatively agree with the LTE and NLTE observational data.

After [Fe/H]$\;=-1$ the [Co/Fe] for N13 CCSNe decreases linearly with metallicity, whereas LC18 CCSNe decreases until [Fe/H]$\;\sim-1.5$, then plateaus until around solar metallicity before decreasing again. Quantitatively, the GCE predictions in Figures \ref{fig:N13_GCE} and \ref{fig:LC18_GCE} that include contributions from SNe Ia also follow these general trends, respectively.

Similar to Mn, Co also has only one stable isotope, $^{59}$Co. This isotope is primarily produced by two radiogenic decay channels, from $^{59}$Ni and $^{59}$Cu, that each dominate at different temperatures depending on the mass of the progenitor \citep{Keegans:2023}. The Co yield has an increasing trend with metallicity in lower mass SN Ia models but remains relatively flat at higher masses. Figures \ref{fig:N13_GCE} \& \ref{fig:LC18_GCE} show that PDf models produce higher [Co/Fe] than all other explosion types. However, although the [Co/Fe] ratio is higher for PDf explosions than others, the absolute yield is low and so they contribute only marginally towards the overall Co inventory when more than one SNe explosion channel is considered. In general, sub-$M_{\text{Ch}}$ SNe Ia lead to higher [Co/Fe] than near-$M_{\text{Ch}}$, at least when PDf models are not considered.   

\subsubsection{Nickel}

Much the same as Mn, the Galactic [Ni/Fe] ratio has also been used as a diagnostic of WD progenitor mass \citep{kirby:2019, kobayashi:mn, blondin:2022, Eitner:23}. Unlike Mn, however, Ni has five stable isotopes $^{58,60,61,62,64}$Ni, with $^{58}$Ni constituting the largest fraction of the element in terms of both SN Ia yields and the solar composition \citep{blondin:2022}. $^{58}$Ni is mostly synthesised during NSE at temperatures $\gtrsim 5$ GK \citep{Brachwitz:2000, blondin:2022, Keegans:2023}. In terms of SN Ia modelling, the metallicity of the progenitor and, by extension, the resulting electron fraction ($Y_e$) is often fine tuned by adjusting the $^{22}$Ne mass fraction in the pre-explosion material. A WD that evolved from a higher metallicity main sequence star will have a higher $^{22}$Ne mass fraction and, since $^{22}$Ne is neutron rich, subsequently a lower $Y_e$ \citep{Hartman:1985}. However, the central density of the WD during the explosive burning of material affects the electron chemical potential \citep{Brachwitz:2000}. This means that WD with higher central densities (i.e. near-$M_{\text{Ch}}$ SNe Ia) have increased Fermi energy and thus enhanced electron-capture rates, which leads to a reduction in $Y_e$ and results in a more efficient production of neutron-rich isotopes \citep{blondin:2022}. The speed of propagation of the flame front during deflagration in near-$M_{\text{Ch}}$ also determines the amount of $^{58}$Ni that is synthesised, as a slower flame speed means that the outer layers of the core have more time to expand before being incinerated. Therefore, slower flame speeds lead to a decrease in the electron-capture rate as a function of radius in the central core, and consequently a reduction in the $^{58}$Ni yield. In summary, the Ni yield i) increases with the density (mass) of the WD progenitor, ii) increases in deflagration models with a faster flame front (e.g. the \citetalias{iwamoto:99} model), iii) has a greater metallicity dependency in sub-$M_{\text{Ch}}$ models, due to the fact that there is no initial deflagration to allow the $Y_e$ to decrease. 

Recently, \cite{Eitner:23} have computed for the first time [Ni/Fe] abundances for a large sample ($>250$) of disk stars including a full treatment of NLTE effects (data provided by P. Eitner, private communication). At low metallicities the NLTE data are higher than LTE measurements, with an average upward shift in [Ni/Fe] of $+0.15$ dex at [Fe/H]$\;=-2$ with respect to the LTE data. This increase in [Ni/Fe] is in good agreement with the sub-[Fe/H]$\;<-1.5$ plateau for this ratio predicted by the LC18 models, however, the N13 models have a highly subsolar [Ni/Fe] ratio at low metallicities. This difference is due to the parameterisation of mixing and fallback in the two CCSN models. In CCSNe, Ni (as $^{58}$Ni) is mostly synthesised during explosive Si inside the innermost regions of the star, and thus the parameterisation of the mixing and fallback determines the amount of Ni that can escape the remnant by being mixed into the ejected material \citep{limongi2018}. For stars in the mass range $13<M<25M_{\odot}$ of the set R yields of LC18 the inner border of the mixed region is calibrated such that [Ni/Fe]$\;=0.2$, whereas for N13 the mixing and fallback parameters are chosen to give the largest [Zn/Fe], which requires a shallower mass cut and leads to a lower [Ni/Fe] yield \citep{Nomoto:2006}. Such a parameterisation was based on LTE Zn measurements, and therefore presents an additional uncertainty when using the N13 yields to produce GCE models.
For metal-poor stars, the available NLTE Zn measurements are typically between $-0.3\lesssim\;$[Zn/Fe]$\;\lesssim0.5$ with one outlier (TYC 891-750-1) having [Zn/Fe] $\sim$ 1.1 dex \citep{Caffau:2023}, while thin disk stars have abundance ratios in the range $-0.2\lesssim\;$[Zn/Fe]$\;\lesssim0.2$ \citep{Sitnova:2022, Caffau:2023}. The N13 CCSN yields vary in the range $-0.8\leq\;$[Zn/Fe]$\;\leq-0.1$, overlapping with the lowest observed values at low metallicity. 

In Fig. \ref{fig:N13_GCE}, there is a steep increase in [Ni/fe] for all near-$M_{\text{Ch}}$ models after [Fe/H]$\;\gtrsim-1.5$, with no clear difference in the rate of increase between the PDf models of \citetalias{fink:2014} and \citetalias{kromer:15} and the DDT models. An exception, however, is \citetalias{iwamoto:99} which increases faster than the others and reaches highly subsolar [Ni/Fe]. The W7 model of \citetalias{iwamoto:99} has been, and indeed still is, adopted by many GCE models, and as such its tendency to grossly overestimate Galactic Ni has been well documented. For all other near-$M_{\text{Ch}}$ models there is good agreement with the observational data, with a slight under or overestimation of [Ni/Fe] relative to the solar value for \citetalias{leung2018}3 and \citetalias{leung2018}9, respectively. 

For the sub-$M_{\text{Ch}}$ models in Fig. \ref{fig:N13_GCE}, the GCE tracks show different behaviours for each of the SN Ia yields: \citetalias{pakmor:12} and \citetalias{gronow:21gce} initially increase, though less rapidly than for near-$M_{\text{Ch}}$, before reaching a plateau of [Ni/Fe]$\;\sim-0.1$ at [Fe/H]$\;\simeq0$; \citetalias{shen:2018}, \citetalias{leung:20}(S), and \citetalias{leung:20}(Y) remain relatively flat until [Fe/H]$\;\sim-0.5$ when they increase; \citetalias{sim:2010} steadily decreases until [Fe/H]$\;\sim-1.0$ and then flattens, following a similar trend as for N13 CCSNe. All sub-$M_{\text{Ch}}$ models in Fig. \ref{fig:N13_GCE} produce subsolar [Ni/Fe] at [Fe/H]$\;=0$.

In Fig. \ref{fig:LC18_GCE}, since the GCE tracks with LC18 have supersolar [Ni/Fe] for [Fe/H]$\;\lesssim-1.5$, an increase in [Ni/Fe] is only seen for \citetalias{iwamoto:99}. Due to the decrease in [Ni/Fe] from LC18 CCSNe after [Fe/H]$\;\sim1.5$, the GCE tracks for \citetalias{seitenzahl:12} and \citetalias{leung2018}3 also decrease from [Ni/Fe]$\;\simeq0.2$ to [Ni/Fe]$\;\simeq0.1$ at [Fe/H]$\;=0$. The remaining near-$M_{\text{Ch}}$ models have GCE tracks that remain relatively flat, resulting in [Ni/Fe]$\;\simeq0.2$ at [Fe/H]$\;=0$.

There is less variation in the chemical evolution patterns of the sub-$M_{\text{Ch}}$ models in Fig. \ref{fig:LC18_GCE}, as Ni production is lower than for near-$M_{\text{Ch}}$ the GCE is more heavily influenced by the sharp decrease in [Ni/Fe] production from LC18 CCSNe in the [Fe/H]$\;\gtrsim-1..5$ regime. Despite this, all sub-$M_{\text{Ch}}$ models in this Figure have good agreement with the observational data after [Fe/H]$\;\gtrsim-1.0$ and lie within 0.1 dex of solar at [Fe/H]$\;=0$.

\subsection{Parameter study of SN Ia yield combinations}\label{sec:combos}

In this Section we test the goodness of fit for GCE models that include contributions from a combination of different SN Ia yields, in order to make predictions about the fraction of sub- and near-$M_{\text{Ch}}$ SNe Ia in the Galaxy. As in the previous Section, we only compare LTE data to GCE models normalised to LTE solar values and NLTE data to NLTE normalised predictions.

Iterating through all sub- and near-${M_{\text{Ch}}}$ yields (except PDf explosions) in Section \ref{sec:yields}, we calculate GCE models using all possible sub- and near-$M_{\text{Ch}}$ SN Ia yield combinations with different fractional contributions of each. For each of the $\sim1200$ GCE models, we perform a $\chi^2$ test in order to quantify the yield combinations and relative fraction of sub/near-$M_{\text{Ch}}$ that best predict the observational data.

For a given GCE model, the $\chi^2$ test is calculated as

\begin{equation}\label{eq:chi}
   \chi^2=\frac{1}{N}\sum^{N}_{i=1} \left( \frac{P_{i}-O_{i}}{\sigma_{O_i}} \right)^2,
\end{equation}
where the sum is taken over the number of observational data points $N$ that have [Fe/H]$\;\geq-1.5$ (such that the calculation is only performed from the time of first SN Ia enrichment). For a given observational measurement with $N=i$, the least-squares is calculated as the difference in [X/Fe] between the observational data $O_i$ and the GCE model prediction $P_i$, where $O_i$ and $P_i$ have the same [Fe/H]. The denominator $\sigma_{O_i}$ is the summation of the uncertainty in [X/Fe] and [Fe/H] that is associated with star $O_{i}$. Since the resolution of the GCE calculation is limited by the number of timesteps, some observational data will have [Fe/H] values that do not correspond exactly to the calculated [Fe/H] of the Galaxy at any given timestep. In this instance, $P_i$ is calculated as $Fn(\text{[Fe/H}_{O_n})$, where $O_n$ is one such observational data point and the function $Fn$ is derived by fitting a cubic interpolation to the GCE prediction. 

In order to place constraints on the relative fraction of sub/near-$M_{\text{Ch}}$ progenitor, for each element and for each set of massive star yields we consider the GCE models with a $\chi^2$ score in the upper $84^{\text{th}}$ percentile. The GCE tracks for models with a combination of sub- and near-$M_{\text{Ch}}$ yields are shown in Appendix \ref{Appendix}. Fig. \ref{fig:chi_frac} shows the $84^{\text{th}}$ percentile probability density function (PDF) for each element as a function of the sub-$M_{\text{Ch}}$ fraction $f_{\text{sub}}$ in the GCE models. The density of the PDF at fraction $f_{\text{sub}}$ indicates the relative number of GCE models in the $84^{\text{th}}$ percentile with $f_{\text{sub}}$ sub-$M_{\text{Ch}}$ yields and $1-f_{\text{sub}}$ near-$M_{\text{Ch}}$ yields. The PDF is normalised such that $\int \text{density}(f_{\text{sub}})df_{\text{sub}}=1$, where the integral is taken over all fractions $f_{\text{sub}}$. The thin blue and black lines indicate PDFs calculated for GCE models fit to LTE data for N13 or LC18 massive star yields, respectively. The thick blue and grey line are calculated using only NLTE data and NLTE GCE normalisation. 

\begin{figure*}[!h]
    \centering
    \includegraphics[scale=0.48]{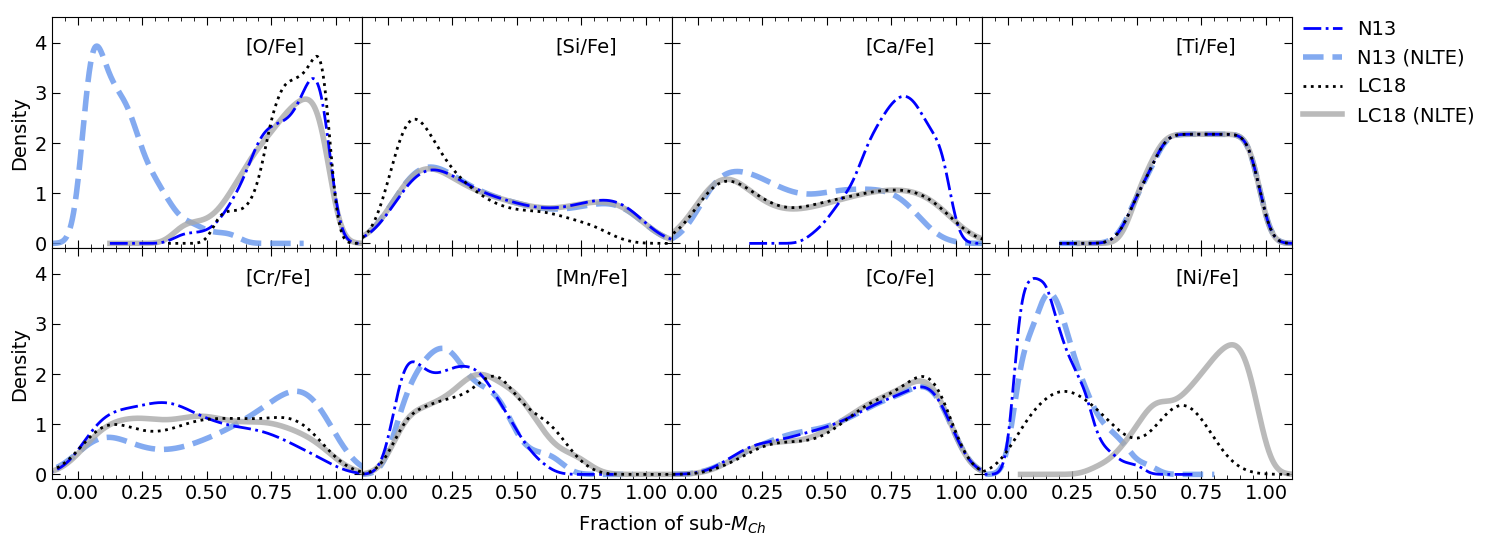}
    \caption{Probability density function of GCE models with a $\chi^2$ score in the $84^{\text{th}}$ percentile, as a function of the fractional contribution from sub-$M_{\text{Ch}}$ SNe Ia in those GCE models (see text for details). The thin dashed blue and dotted black lines include only LTE observational data for each of the GCE models with N13 and LC18 massive star yields, respectively. The thick solid blue and grey lines are calculated using only NLTE data. The results should be interpreted in the context that, for some elements, the absence of well-fitting GCE models may compromise the reliability of the observed patterns.}
    \label{fig:chi_frac}
\end{figure*}

\begin{figure*}[!h]
    \centering
    \includegraphics[width=1.0\textwidth]{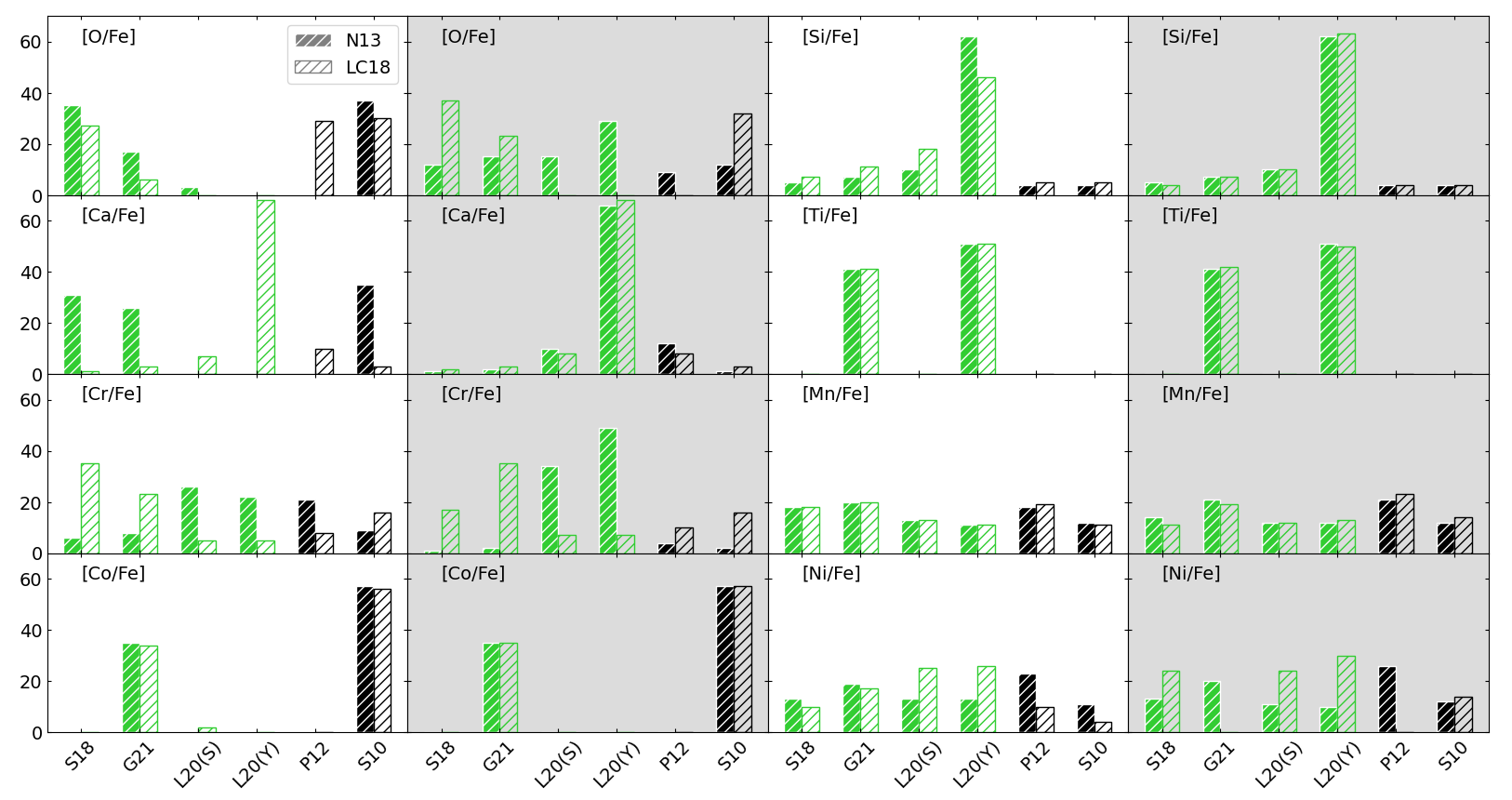}
    \includegraphics[width=1.0\textwidth]{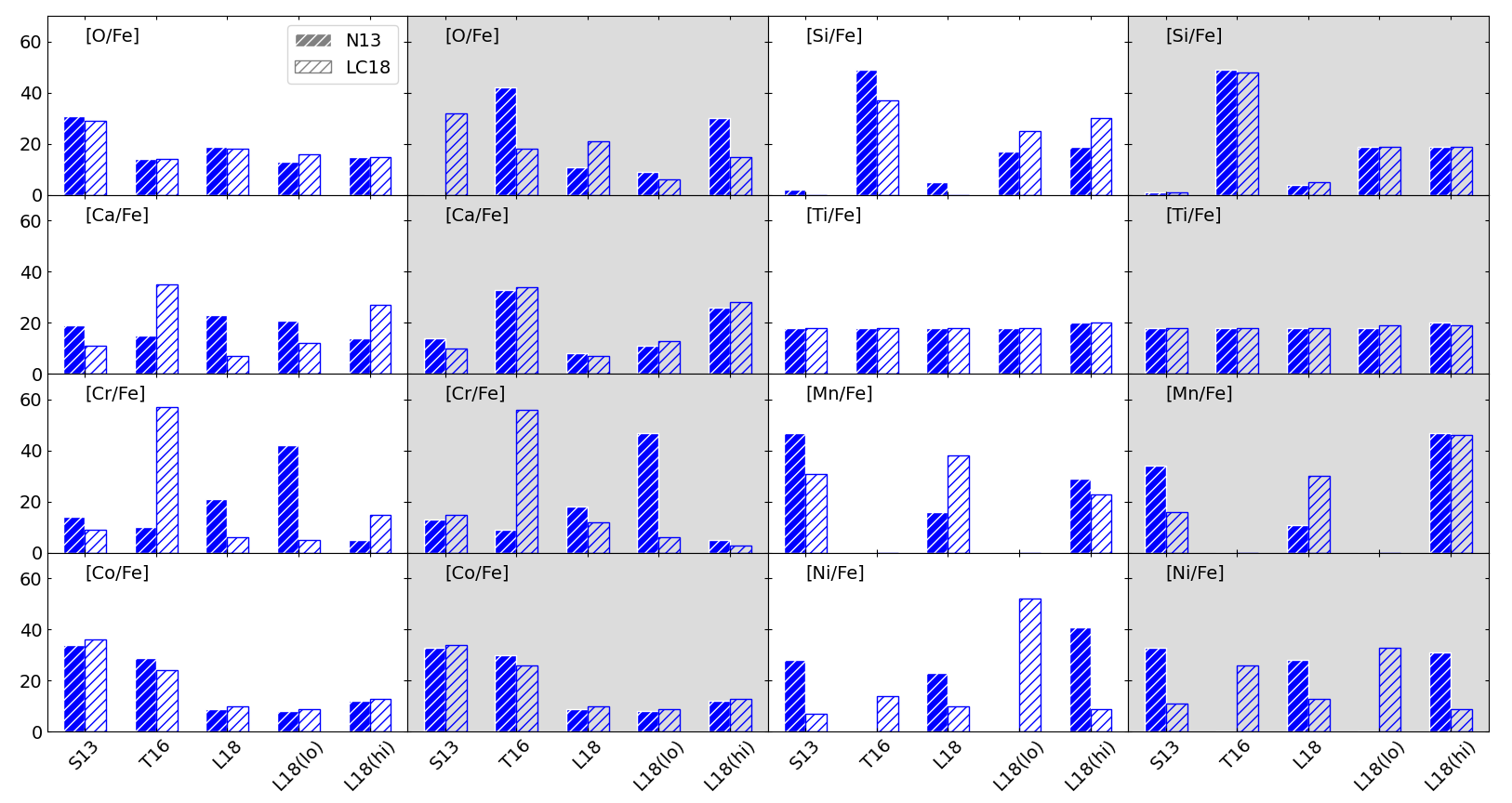}
    \caption{Counts of sub-$M_{\text{Ch}}$ (top panel) and near-$M_{\text{Ch}}$ (bottom panel) SN Ia yields that are used by N13 and LC18 GCE models with a $\chi^2$ score in the $84^{\text{th}}$ percentile. The bar colours indicate the explosion mechanism, as in Figures \ref{fig:N13_GCE} and \ref{fig:LC18_GCE}. Shaded and unshaded panels are for calculations using NLTE and LTE observational datasets, respectively.}
    \label{fig:frac_quartiles}
\end{figure*}

Further to examining the spread of $f_{\text{sub}}$ in the best fit GCE models, for each element we investigate the number of GCE models in the $84^{\text{th}}$ percentile based on the specific choice of SN Ia yields. Fig. \ref{fig:frac_quartiles} shows a histogram of the number of GCE models in the $84^{\text{th}}$ percentile based on the choice of sub-$M_{\text{Ch}}$ (top panel) and near-$M_{\text{Ch}}$ yields (bottom panel). By examining Figures \ref{fig:chi_frac} and \ref{fig:frac_quartiles} in tandem, we can see the sub/near-$M_{\text{Ch}}$ ratios and SN Ia models that provide the best fit to the observational data and how much the choice of massive star yields can affect the constraining capabilities of such GCE tests. 

As discussed in Section \ref{sec:GCE_vs_obs}, several of the Fe-peak element data trends cannot be well reproduced by any of the GCE models over the relevant metallicity range in the Galactic disk, mostly due to limitations regarding current stellar nucleosynthesis yields. Therefore, for elements such as Co and Ti, the best-fitting GCE models as described by the PDFs in Fig. \ref{fig:chi_frac} cannot be used yet as robust constraints for SN Ia progenitors in the Galaxy. Instead, they serve to strengthen the argument that until further improvements in calculations of massive star explosions are made, GCE models cannot be safely used to constrain SN Ia progenitors. In the following discussion we are careful to highlight such weaknesses in regards to using GCE as a diagnostic tool of progenitor mass.  

\textit{Oxygen:} The difference between the best and worst fitting GCE models in Figures \ref{fig:N13_GCE_chi} \& \ref{fig:LC18_GCE_chi} is substantially smaller than the spread in the [O/Fe] data. This is to be expected, since SNe Ia do not produce O so there is very little variation in the O and Fe yields between the different SN Ia models. Interestingly, for both N13 and LC18, the LTE data are best fit by models with a high $f_{\text{sub}}$. However, the $f_{\text{sub}}$ is shifted downwards -- significantly for the former -- when comparing the predictions to NLTE data. Although a good test for the calibration of GCE models to reproduce the chemical behaviour of the Milky Way disks, O clearly cannot be used to probe the distribution of different SN Ia progenitors.     

\textit{Silicon and Calcium:} In general, there is a slight preference for lower $f_{\text{sub}}$ among the best-fitting GCE models to the Si data. In particular, Fig. \ref{fig:frac_quartiles} shows that the \citetalias{leung:20}(Y) model is commonly used in the best fitting models for both CCSN yields sets when comparing to both LTE and NLTE observations. This SN Ia explosion produces the lowest amount of Si, so can better compensate for the slight overproduction of [Si/Fe] with respect to the data from the massive stars in our GCE models. For Ca, if either NLTE corrections are included or LC18 CCSNe are used, there are a large range of $f_{\text{sub}}$ values (between $\sim0.1-0.9$) that fit the data well. However, the LC18 models do not well fit the LTE or NLTE [Ca/Fe] observational data below solar metallicity, so we advise caution when interpreting the results for these models in Fig. \ref{fig:chi_frac}. For N13 yields with LTE approximation, there is a high density of $f_{\text{sub}}\approx0.75$ models that best fit the data. Recently, \cite{Cavichia:2024} calculated 180 chemical evolution models with LC18 CCSN yields and with different DTD and SN Ia yields to compare to $\alpha$-element observations of the Milky Way's disk. They find the best fit to the data are obtained using a DTD for a single-degenerate scenario and with SN Ia yields from \cite{leung2018}, with $75\%$ of their best-fitting models including yields from near-$M_{\text{Ch}}$. We note, however, that their GCE models do not include different combinations of sub- and near-$M_{\text{Ch}}$ yields, and that their chosen observational datasets appear to include a mix of LTE and NLTE data. Based on the calculations presented in this work, both of these features could impact their results.

\textit{Titanium:} In Fig. \ref{fig:chi_frac} the PDF for [Ti/Fe] is identical for both N13 and LC18 GCE models, regardless of whether comparisons are made to LTE or NLTE data. Fig. \ref{fig:frac_quartiles} (bottom panel) shows that there is an almost even distribution of the adopted near-$M_{\text{Ch}}$ yields in the best-fitting GCE models. The PDF shows that nearly all GCE models in the $84^{\text{th}}$ percentile have a greater than $0.5$ fraction of sub-$M_{\text{Ch}}$ yields which, when examining Fig. \ref{fig:frac_quartiles}, are exclusively from the \citetalias{leung:20}\_M10\_05(Y) and \citetalias{gronow:21gce} sub-$M_{\text{Ch}}$ models. Since all GCE models in this percentile use only one of two sub-$M_{\text{Ch}}$ yields, the peak of the PDF is relatively flat and extends between $0.6<f_{\text{sub}}<0.9$. We remind that the \citetalias{gronow:21gce} and \citetalias{leung:20}\_M11\_05(Y) GCE models produce distinct tracks compared to the other PDt or DD explosions. Furthermore, the Y explosion pattern is the only example of a non-spherical explosion geometry used in this analysis, so we must be careful not to extrapolate a conclusion about WD progenitor mass if the geometry could be partly or wholly accountable.

Our results for Ti agree with \cite{palla:2021}, who reason that a significant contribution from DD SN Ia explosions with aspherical He detonation (i.e. non S-type) are required to match solar Ti. We also find that GCE models using \citetalias{gronow:21gce} lead to similar $\chi^2$ values to those of \citetalias{leung:20}\_M11\_05(Y). The reason for this is unclear, however, in \cite{gronow:21gce} the authors note that increasing the contribution of \citetalias{gronow:21gce} in their GCE calculations so as to match solar Mn, would lead to an overestimation of the Ti yield. Although our statistical test concludes that Ti production favours a dominant $f_{\text{sub}}$ in the Galaxy, we remind than none of the GCE models are able to well reproduce the data in the low metallicity regime. A higher production of Ti in CCSNe, or a significant downwards shift in low-metallicity [Ti/Fe] NLTE data, so as to reconcile the current disagreement between theory and observation is therefore required before the constraining potential of this element can be validated.

\textit{Chromium:} We remind that the PDFs for NLTE Cr are calculated using only three data points. Regardless, the NLTE calculations give consistent results for Cr I and Cr II in this metallicity range and are well fit by the LC18 models. The PDF for [Cr/Fe] has subtle differences depending on the choice of massive star yields. For N13, the PDF is weighted towards lower values of $f_{\text{sub}}$ with a small peak at 0.3, although the tail of the distribution extends all the way to $f_{\text{sub}}=1$. The distribution for LC18 is wider, and there are almost equal numbers of models for all values of $f_{\text{sub}}$. A wide distribution indicates that there are a range of $f_{\text{sub}}$ that fit the observational data to an almost equal degree. The LC18 models have subsolar [Cr/Fe] at the onset of GCE thus, by virtue of being the models that produce the highest amount of Cr, the \citetalias{shen:2018}, \citetalias{gronow:21gce} and \citetalias{townsley:16} yields are the most common among GCE models in the $84^{\text{th}}$ percentile. On the other hand, the N13 models have a supersolar [Cr/Fe] plateau at low metallicities, and as such are best fit by SN Ia models that produce the least amount of Cr, chiefly the low-density near-$M_{\text{Ch}}$ model of \cite{leung2018}. These results indicate that from a GCE perspective Cr is a relatively weak identifier of the progenitor mass of SNe Ia, since the same general trends are seen in sub- and near-$M_{\text{Ch}}$ models.

\textit{Manganese:} We remind that NLTE corrections to Mn data are strictly positive at low metallicities, resulting in the NLTE data having a near-solar plateau at [Fe/H]$>-2$. At present, the Mn yields from CCSN models are too low to account for the near-solar plateau in the NLTE observations, resulting in the GCE models having a good fit to the data only after $\sim-0.5\;$ [Fe/H]. Since both the LC18 and N13 models have sub-solar [Mn/Fe] prior to SN Ia enrichment, GCE models that include large fractions of SNe Ia with higher Mn yields (i.e. near-$M_{\text{Ch}}$) are necessary to fit the NLTE data. However, at least some fraction of sub-$M_{\text{Ch}}$ SNe Ia are required to prevent the overproduction of [Mn/Fe] at super solar metallicities. For both N13 and LC18 the PDF for LTE and NLTE data are very similar, with peaks at $f_{\text{sub}}\approx0.2$ and 0.4, respectively. The higher $f_{\text{sub}}$ for LC18 is due to the comparatively higher Mn yields for the CCSN models at and around solar metallicity. For comparison to other LTE studies, the peak fraction of $f_{\text{sub}}\simeq0.2$ for N13 GCE models agrees with \cite{kobayashi:mn}, who conclude that sub-$M_{\text{Ch}}$ must contribute up to $\sim25\%$ of SN Ia chemical enrichment when using similar massive star yields to N13. The $f_{\text{sub}}$ distribution for LC18 are similar to the results of \cite{seitenzahl:Mn}, who find that a 50-50 mix of sub- and near-$M_{\text{Ch}}$ channels provide a best fit to the data. In regards to the NLTE results, \cite{palla:2021} find that for their GCE model with standard CCSNe from \cite{kobayashi:2006} (similar to the N13 yields) the best-fitting model to the MnII LTE and Mn NLTE data has a $25\%$ contribution from sub-$M_{\text{Ch}}$. This is in good agreement with the NLTE N13 results in Fig \ref{fig:chi_frac}, which has a peak in the distribution at $f_{\text{sub}}\approx0.2$. In \cite{eitner:2020}, the Mn yields from \cite{Woosley:1995} CCSNe are increased by $50\%$ to provide a better agreement with the solar plateau of low-metallicity NLTE data for [Mn/Fe], in which case the best fit GCE model has a $75\%$ contribution from sub-$M_{\text{Ch}}$ SNe Ia. The authors also claim that the same results can be obtained using the standard CCSN yields from \cite{Kobayashi:2011}, however, these claims are not substantiated by the results in this work using the N13 yields of the same group, which have a sub-solar plateau of [Mn/Fe]$\;=-0.5$ at [Fe/H]$\;<-1.5$.

Regardless of the CCSN yields used, it is interesting that at least some fraction of sub-$M_{\text{Ch}}$ are still necessary to better fit the NLTE [Mn/Fe] trend. Furthermore, Fig \ref{fig:frac_quartiles} shows that the best-fitting GCE models contain an almost equal spread of the different sub-$M_{\text{Ch}}$ SN Ia yields, and three of the five near-$M_{\text{Ch}}$ yields. Therefore, the PDF was not skewed by one or two good fitting yield sets (as for Ti and Co), and is proof that [Mn/Fe] is a robust ratio to use to probe SN Ia progenitor mass. It is unquestionable that if future Mn yields from CCSNe did increase so as to better fit the low metallicity plateau of [Mn/Fe], then the accompanying GCE models would predict a higher $f_{\text{sub}}$ -- likely in agreement with the results of \cite{eitner:2020} and \cite{palla:2021}. However, until this correction is made, we refrain from applying nonphysical alterations to the yields since Mn is not isolated from the production of other Fe-peak elements during NSE, and it is not yet clear if/how these other elements would also be effected.

In Section \ref{sec:results}, it was found that GCE models with PDf SN Ia yields lead to supersolar [Mn/Fe]. However, these explosions eject too little $^{56}$Ni to be considered `normal' SNe Ia, and are instead the leading candidate to explain the subluminous SN Iax subclass \citep[\emph{c.f.}][]{Karambelkar:2022}. Binary population synthesis models have shown that SNe Iax can have delay times much shorter than typical SNe Ia, leading to a prompt enhancement of [Mn/Fe] in the Galaxy. To test whether a prompt contribution from .Iax can help correct for the [Mn/Fe] deficit predicted by GCE models when compared to NLTE observations, we run additional GCE models that take into account also a contribution from SNe Iax. In these models we assume that SNe Iax explode at a rate of $20\%$ of normal SNe Ia, such that the total SNe Ia rate is $1.56\times10^{-3}\;M_{\odot}^{-1}$, and their delay time distribution peaks sharply at $\sim100$ Myr (see, e.g. \cite{liu:2015}). Since low-luminosity SNe Iax make up the bulk of the total SNe Iax rate \citep{srivastav:2022}, we use the yields from the \citetalias{kromer:15} model, which is shown to have good agreement to the subluminous SN 2008ha. Our statistical test confirms that the $f_{\text{sub}}$ is identical for GCE models with and without a contribution from SNe Iax. Therefore, we claim that SN Iax events are both too rare and eject too little Mn to affect the evolution of [Mn/Fe], even at earlier Galactic times. 

\textit{Cobalt:} For [Co/Fe], the distribution is again near identical for N13 and LC18, with a strong correlation between the goodness of fit of the GCE model and a higher $f_{\text{sub}}$. Fig. \ref{fig:frac_quartiles} shows that the majority of sub-$M_{\text{Ch}}$ yields in the $84^{\text{th}}$ percentile are \citetalias{gronow:21gce} and \citetalias{sim:2010}, with a more varied distribution for near-$M_{\text{Ch}}$ yields. However, none of the GCE models provide a good fit to the data across the entire metallicity range, thus effectively ruling out at present the use of Co as a strong constraint of the SN Ia population.  

\textit{Nickel:} The largest difference in $f_{\text{sub}}$ between N13 and LC18 GCE models is for [Ni/Fe]. Since [Ni/Fe] is supersolar for LC18 at [Fe/H]$\;<-1.5$ (we remind this is due to the choice of mixing and fallback which determines the position of the mass cut), a higher $f_{\text{sub}}$ is favoured in the GCE models since sub-$M_{\text{Ch}}$ SNe Ia have lower [Ni/Fe] than near-$M_{\text{Ch}}$. Conversely, the [Ni/Fe] plateau before SN Ia enrichment in the N13 GCE models is sub-solar, meaning the GCE predictions fail to align well with both LTE and NLTE observational data below approximately $-0.25\;$ [Fe/H]. Therefore, a greater fraction of near-$M_{\text{Ch}}$ SNe Ia are required to compensate for the reduced Ni production from the N13 CCSNe. In Fig. \ref{fig:chi_frac}, the N13 probability density distributions for LTE and NLTE have well-defined peaks at $f_{\text{sub}}\simeq0.1$ and $f_{\text{sub}}\simeq0.2$, respectively. For LC18, the LTE distribution has two peaks at $f_{\text{sub}}\simeq0.2$ and 0.65; the first peak is comprised mostly of GCE models using the near-$M_{\text{sub}}$ yields from the \citetalias{leung2018}3 model, whereas the models that form the second peak contain the other near-$M_{\text{sub}}$ yields in combination with different sub-$M_{\text{sub}}$ yields. Regarding the NLTE distribution, the LC18 distribution has a peak at $f_{\text{sub}}\simeq0.85$, and contains a very small density of models with $f_{\text{sub}}<0.5$. We remind that when NLTE corrections are taken into account, the average [Ni/Fe] value increases across all metallicities, especially in the [Fe/H]$\;<-1$ regime. The flatness of the [Ni/Fe] versus metallicity trend for the NLTE data are best fit by models including a high number of sub-$M_{\text{sub}}$ SNe Ia, otherwise the near-$M_{\text{sub}}$ SN Ia contribution leads to overestimates of [Ni/Fe] in the super-solar regime. The well defined peak for the NLTE PDFs in Fig. \ref{fig:chi_frac} indicates that $f_{\text{sub}}$ is more important for reproducing the observational data than the specific SN Ia yields. This is confirmed in Fig. \ref{fig:frac_quartiles}, since no one sub- or near-$M_{\text{Ch}}$ yield dominates in the best-fitting GCE models. 

The results of \cite{Eitner:23} provide an interesting comparison with our work, since \cite{Eitner:23} also focused their analysis on the impact of SN Ia yields on GCE using the \texttt{OMEGA+} code. There are indeed important differences. In this work we make the simplification that all SNe Ia explode according to a power law delay-time-distribution (DTD), as expected for double degenerate systems of WD-WD mergers. In \cite{Eitner:23}, the chemical evolution of the [Ni/Fe] abundance ratio in the Galactic disk is modelled for four different SN Ia progenitor systems, each with unique DTDs derived using a population synthesis code. As in this work, in \cite{Eitner:23} a statistical test is used to determine the goodness of fit of the models to NLTE observational data, wherein it is found that models with $f_{\text{sub}}\sim 80\%$ provide the best fit. This value is in excellent agreement with the $f_{\text{sub}}$ values found in this work for the NLTE [Ni/Fe] distribution for GCE models with CCSNe from LC18. \cite{Eitner:23}, however, only use the [Ni/Fe] ratio in their calculation of $f_{\text{sub}}$. In this work we instead consider most of the Fe-peak elements in the analysis, and the $\alpha$-elements O, Si and Ca. These elements are produced in a similar physical environment and ejected together by both CCSNe and SNe Ia. Their stellar production cannot be disentangled, and provides more additional constraints when GCE models are compared with observations. In the LC18 CCSN models \citep[used as a reference by][in their conclusions]{Eitner:23} the position of the inner border of the mixing and fallback prescription that determines the [Ni/Fe] ejected during the explosion is fixed such that [Ni/Fe]$\;=0.2$. Furthermore, the Fe yield is ejected predominantly as $^{56}$Ni made during complete Si burning, where the ejecta contains a fixed amount of this radioisotope in order to reproduce the main properties of the observed CCSN light curve. This indeed may affect the results to constrain $f_{\text{sub}}$ using only the [Ni/Fe] ratio obtained from GCE models, since the yield of both of these elements in the ejected material of CCSNe are chosen `ad hoc' rather than through purely physical means. Therefore, for GCE models using CCSN yields by LC18 we advise that the values of $f_{\text{sub}}$ derived using the [Ni/Fe] ratio should be interpreted with caution. 

\section{Discussion and conclusion}\label{sec:conclusion}

In this work, we investigated the GCE of $\alpha$- (O, Si, and Ca) and Fe-peak elements (Ti, Cr, Mn, Co, and Ni) in the Milk Way's disk using yields from four different SN Ia explosion mechanisms and two sets of massive star yields: N13 (non-rotating) and LC18 (rotating). We compare our GCE predictions to LTE and NLTE spectroscopic measurements in stars and describe the results by analysing the nucleosynthesis conditions that are required to produce each element during the explosion. In particular, we consider how the predicted evolution depends on the mass of the WD progenitor and the choice of massive star yields. We calculate GCE models that include contributions from both sub- and near-$M_{\text{Ch}}$ SNe Ia, including a suite of GCE models that take into account a 20$\%$ contribution from the faint Iax subclass. For each element, a statistical test is used to score the models based on the goodness of fit to the stellar data, and a probability density estimation is made over the fraction of sub-$M_{\text{Ch}}$ SNe Ia $(f_{\text{sub}})$ for N13 and LC18 GCE models with a $\chi^2$ score in the $84^{\text{th}}$ percentile. These are cross-compared with the distribution of SN Ia yields that are used in the best-fitting GCE models to help constrain the relative fraction of sub- and near-$M_{\text{Ch}}$ SNe Ia in the Galaxy. 

Our results show that the present uncertainties and large variations affecting the yields of the Fe-peak elements in CCSN yields do not allow the use of GCE simulations alone to constrain the populations of SN Ia progenitors in the Milky Way disk. Our main results are summarised below.  

\begin{itemize}

    \item{The O and Si data (LTE and NLTE) are well fit by both the N13 and LC18 GCE models, whereas the [Ca/Fe] data can only be fit across the entire metallicity range using N13 yields. Although Si and Ca are made in small amounts by SNe Ia, the variation between yields of different progenitor mass are not significant enough to constrain contribution to GCE from different types of SNe Ia.}
 
    \item{Of the Fe-peak elements, we confirm that Ti and Co data are not well fit by any combination of CCSN and SN Ia yields, and therefore their capability to constrain the nature of SN Ia progenitor is limited. On the other hand, a combination of sub- and near-$M_{\text{Ch}}$ yields can reproduce the majority of Cr and Mn data after [Fe/H]$\;>-1$. However, at lower metallicities, the choice of massive star yields is significant for the GCE of Cr. At [Fe/H]$\;\lesssim-1.5$ the non-rotating massive star yields of N13 are $\sim0.4$ dex higher for [Cr/Fe] than for the rotating massive star yields of LC18.}

    \item{The GCE of [Mn/Fe] and [Ni/Fe] after [Fe/H]$\;\approx-1.5$ are sensitive to the relative fraction of SNe Ia in the Galaxy from the sub- and near-$M_{\text{Ch}}$ WD progenitor channels. However, the inter-channel explosion mechanisms (i.e. whether DDT or PDF yields are used for near-$M_{\text{Ch}}$, or DD or PDt for sub-$M_{\text{Ch}}$) do not significantly change the GCE of either element.}
    
    \item{We have shown that the $f_{\text{sub}}$ that best fit the [Mn/Fe] spectroscopic data depends on the choice of massive star yields and, additionally for [Ni/Fe], whether NLTE corrections are included. Therefore, we argue that future GCE studies must include a complete NLTE treatment (i.e. both data and GCE normalisation) in order to draw relevant conclusions about the relative GCE contribution from the different types of SNe Ia. Furthermore, the constraining potential of GCE studies is severely limited by the current disagreement regarding the Mn and Ni yields from different CCSN yield sets.}

    \item{For [Mn/Fe], the best fitting GCE models to the NLTE data have $f_{\text{sub}}\approx0.2$ (N13) and $f_{\text{sub}}\approx0.4$ (LC18), however, even with these fractions the corresponding GCE models do not well fit the data. For NLTE [Ni/Fe], the models that scored the highest in the goodness of fit test had predominantly $f_{\text{sub}}\approx0.2$ (N13) and $f_{\text{sub}}\approx0.85$ (LC18), though only the latter can fit the data across the entire metallicity range. We advise that in particular the $f_{\text{sub}}$ values calculated using the [Mn/Fe] ratio are likely to increase if future CCSN yields are able to better reproduce the flat trend of [Mn/Fe] NLTE data at low metallicities.}
    
    \item{Despite the fact they produce supersolar [Mn/Fe], an additional prompt enrichment of SNe Iax results in a negligible increase in [Mn/Fe] at low metallicities. This is because their absolute yields are too low and they are too few in number to counterbalance the [Mn/Fe] deficit from massive stars compared to NLTE abundance measurements in the [Fe/H]$\;\lesssim-1.5$ regime.}
    
    \item{For [Ti/Fe], despite the systematic underproduction of this ratio for the GCE models at lower metallicities, both the \citetalias{leung:20}\_M11\_05(Y) model with ring He detonation and \citetalias{gronow:21gce} model lead to predicted values within 0.1 dex at solar-like metallicities. In comparison, the \citetalias{leung:20}\_M10\_05 model with spherical He detonation leads to [Ti/Fe]$\;\simeq-0.3$ at [Fe/H]$\;=0$. In terms of the other Fe-peak elements, the \citetalias{leung:20}\_M11\_05(Y) model is unremarkable from other DD and PDt models, which could suggest that at least some fraction of sub-$M_{\text{Ch}}$ SNe Ia have an aspherical He detonation pattern; this notion was also proposed by \cite{palla:2021}.}
    
\end{itemize}

\begin{acknowledgements}

We thank P. Eitner for providing the [Ni/Fe] NLTE stellar abundances from Eitner et al. (2023). We extend our thanks also to Lorenzo Roberti for fruitful discussions regarding CCSN nucleosynthesis and the anonymous referee for the valuable comments and suggestions for this work. This work made use of the Heidelberg Supernova Model Archive (HESMA), \url{https://hesma.h-its.org}. We are grateful to Ashley Ruiter for providing SN Ia delay-time data to compare with our fiducial rate. TT, MP, and BKG acknowledge the support of STFC (through the University of Hull's Consolidated Grant ST/R000840/1) and ongoing access to {\tt viper}, the University of Hull High Performance Computing Facility. BC, MP, and BKG thank the National Science Foundation (NSF, USA) under grant No. PHY-1430152 (JINA Center for the Evolution of the Elements). BKG and MP acknowledge the support of the European Union’s Horizon 2020 research and innovation programme (ChETEC-INFRA – Project no. 101008324). BC and MP acknowledge the support from the Lend\"ulet Program LP2023-10 of the Hungarian Academy of Sciences (Hungary) and the ERC Consolidator Grant (Hungary) funding scheme (Project RADIOSTAR, G.A. n. 724560). MP acknowledges support from the NKFI via K-project 138031 (Hungary).
\end{acknowledgements}

\section*{Data availability}

The GCE model data generated for this article will be shared on reasonable
request to the corresponding author. Data from stellar observations used for comparison in this article 
can be found in the NuPyCEE package  \url{https://github.com/NuGrid/NuPyCEE/tree/master/stellab_data/milky_way_data}.


\bibliography{sample631}

\begin{thebibliography}{119}
\expandafter\ifx\csname natexlab\endcsname\relax\def\natexlab#1{#1}\fi

\bibitem[{{Adibekyan} {et~al.}(2012){Adibekyan}, {Sousa}, {Santos}, {Delgado Mena}, {Gonz{\'a}lez Hern{\'a}ndez}, {Israelian}, {Mayor}, \& {Khachatryan}}]{adibekyan:2012}
{Adibekyan}, V.~Z., {Sousa}, S.~G., {Santos}, N.~C., {et~al.} 2012, \aap, 545, A32

\bibitem[{{Arnett}(1969)}]{Arnett:1969}
{Arnett}, W.~D. 1969, \apj, 157, 1369

\bibitem[{{Asplund} {et~al.}(2021){Asplund}, {Amarsi}, \& {Grevesse}}]{Asplund:2021}
{Asplund}, M., {Amarsi}, A.~M., \& {Grevesse}, N. 2021, \aap, 653, A141

\bibitem[{{Battino} {et~al.}(2020){Battino}, {Pignatari}, {Travaglio}, {Lederer-Woods}, {Denissenkov}, {Herwig}, {Thielemann}, \& {Rauscher}}]{Battino:2020}
{Battino}, U., {Pignatari}, M., {Travaglio}, C., {et~al.} 2020, \mnras, 497, 4981

\bibitem[{{Battistini} \& {Bensby}(2015)}]{Battistini:2015}
{Battistini}, C. \& {Bensby}, T. 2015, \aap, 577, A9

\bibitem[{{Belczynski} {et~al.}(2008){Belczynski}, {Kalogera}, {Rasio}, {Taam}, {Zezas}, {Bulik}, {Maccarone}, \& {Ivanova}}]{Belczynski:2008}
{Belczynski}, K., {Kalogera}, V., {Rasio}, F.~A., {et~al.} 2008, \apjs, 174, 223

\bibitem[{{Bensby} {et~al.}(2005){Bensby}, {Feltzing}, {Lundstr{\"o}m}, \& {Ilyin}}]{Bensby:2005}
{Bensby}, T., {Feltzing}, S., {Lundstr{\"o}m}, I., \& {Ilyin}, I. 2005, \aap, 433, 185

\bibitem[{{Bergemann}(2011)}]{Bergemann:2011}
{Bergemann}, M. 2011, \mnras, 413, 2184

\bibitem[{{Bergemann} \& {Cescutti}(2010)}]{Bergemann:2010}
{Bergemann}, M. \& {Cescutti}, G. 2010, \aap, 522, A9

\bibitem[{{Bergemann} {et~al.}(2019){Bergemann}, {Gallagher}, {Eitner}, {Bautista}, {Collet}, {Yakovleva}, {Mayriedl}, {Plez}, {Carlsson}, {Leenaarts}, {Belyaev}, \& {Hansen}}]{bergemann:2019}
{Bergemann}, M., {Gallagher}, A.~J., {Eitner}, P., {et~al.} 2019, \aap, 631, A80

\bibitem[{{Bertran de Lis} {et~al.}(2015){Bertran de Lis}, {Delgado Mena}, {Adibekyan}, {Santos}, \& {Sousa}}]{Bertran:2015}
{Bertran de Lis}, S., {Delgado Mena}, E., {Adibekyan}, V.~Z., {Santos}, N.~C., \& {Sousa}, S.~G. 2015, \aap, 576, A89

\bibitem[{{Blondin} {et~al.}(2022){Blondin}, {Bravo}, {Timmes}, {Dessart}, \& {Hillier}}]{blondin:2022}
{Blondin}, S., {Bravo}, E., {Timmes}, F.~X., {Dessart}, L., \& {Hillier}, D.~J. 2022, \aap, 660, A96

\bibitem[{{Brachwitz} {et~al.}(2000){Brachwitz}, {Dean}, {Hix}, {Iwamoto}, {Langanke}, {Mart{\'\i}nez-Pinedo}, {Nomoto}, {Strayer}, {Thielemann}, \& {Umeda}}]{Brachwitz:2000}
{Brachwitz}, F., {Dean}, D.~J., {Hix}, W.~R., {et~al.} 2000, \apj, 536, 934

\bibitem[{{Caffau} {et~al.}(2023){Caffau}, {Lombardo}, {Mashonkina}, {Sitnova}, {Bonifacio}, {Matas Pinto}, {Kordopatis}, {Sestito}, {Aguado}, {Salvadori}, {Spite}, {Fran{\c{c}}ois}, {Sbordone}, {Mucciarelli}, {Martin}, {Fern{\'a}ndez-Alvar}, \& {Gonz{\'a}lez Hern{\'a}ndez}}]{Caffau:2023}
{Caffau}, E., {Lombardo}, L., {Mashonkina}, L., {et~al.} 2023, \mnras, 518, 3796

\bibitem[{{Cavichia} {et~al.}(2024){Cavichia}, {Moll{\'a}}, {Baz{\'a}n}, {Castrillo}, {Galbany}, {Mill{\'a}n-Irigoyen}, {Ascasibar}, {D{\'\i}az}, \& {Monteiro}}]{Cavichia:2024}
{Cavichia}, O., {Moll{\'a}}, M., {Baz{\'a}n}, J.~J., {et~al.} 2024, \mnras, 532, 2331

\bibitem[{{Cescutti} {et~al.}(2008){Cescutti}, {Matteucci}, {Lanfranchi}, \& {McWilliam}}]{Cescutti:2008}
{Cescutti}, G., {Matteucci}, F., {Lanfranchi}, G.~A., \& {McWilliam}, A. 2008, \aap, 491, 401

\bibitem[{{Chiappini} {et~al.}(1997){Chiappini}, {Matteucci}, \& {Gratton}}]{chiappini:1997}
{Chiappini}, C., {Matteucci}, F., \& {Gratton}, R. 1997, \apj, 477, 765

\bibitem[{{Chieffi} {et~al.}(1998){Chieffi}, {Limongi}, \& {Straniero}}]{Chieffi:1998}
{Chieffi}, A., {Limongi}, M., \& {Straniero}, O. 1998, \apj, 502, 737

\bibitem[{{C{\^o}t{\'e}} {et~al.}(2017){C{\^o}t{\'e}}, {O'Shea}, {Ritter}, {Herwig}, \& {Venn}}]{cote17}
{C{\^o}t{\'e}}, B., {O'Shea}, B.~W., {Ritter}, C., {Herwig}, F., \& {Venn}, K.~A. 2017, \apj, 835, 128

\bibitem[{{C{\^o}t{\'e}} {et~al.}(2016){C{\^o}t{\'e}}, {Ritter}, {O'Shea}, {Herwig}, {Pignatari}, {Jones}, \& {Fryer}}]{cote:2016}
{C{\^o}t{\'e}}, B., {Ritter}, C., {O'Shea}, B.~W., {et~al.} 2016, \apj, 824, 82

\bibitem[{{C{\^o}t{\'e}} {et~al.}(2018){C{\^o}t{\'e}}, {Silvia}, {O'Shea}, {Smith}, \& {Wise}}]{cote18}
{C{\^o}t{\'e}}, B., {Silvia}, D.~W., {O'Shea}, B.~W., {Smith}, B., \& {Wise}, J.~H. 2018, \apj, 859, 67

\bibitem[{{Cristallo} {et~al.}(2015){Cristallo}, {Straniero}, {Piersanti}, \& {Gobrecht}}]{cristallo:2015}
{Cristallo}, S., {Straniero}, O., {Piersanti}, L., \& {Gobrecht}, D. 2015, \apjs, 219, 40

\bibitem[{{de los Reyes} {et~al.}(2020){de los Reyes}, {Kirby}, {Seitenzahl}, \& {Shen}}]{Delosreyes:2020}
{de los Reyes}, M. A.~C., {Kirby}, E.~N., {Seitenzahl}, I.~R., \& {Shen}, K.~J. 2020, \apj, 891, 85

\bibitem[{{Denissenkov} {et~al.}(2017){Denissenkov}, {Herwig}, {Battino}, {Ritter}, {Pignatari}, {Jones}, \& {Paxton}}]{Denissenkov:2017}
{Denissenkov}, P.~A., {Herwig}, F., {Battino}, U., {et~al.} 2017, \apjl, 834, L10

\bibitem[{{Denissenkov} {et~al.}(2013){Denissenkov}, {Herwig}, {Truran}, \& {Paxton}}]{Denissenkov:2013}
{Denissenkov}, P.~A., {Herwig}, F., {Truran}, J.~W., \& {Paxton}, B. 2013, \apj, 772, 37

\bibitem[{{Eitner} {et~al.}(2020){Eitner}, {Bergemann}, {Hansen}, {Cescutti}, {Seitenzahl}, {Larsen}, \& {Plez}}]{eitner:2020}
{Eitner}, P., {Bergemann}, M., {Hansen}, C.~J., {et~al.} 2020, \aap, 635, A38

\bibitem[{{Eitner} {et~al.}(2023){Eitner}, {Bergemann}, {Ruiter}, {Avril}, {Seitenzahl}, {Gent}, \& {C{\^o}t{\'e}}}]{Eitner:23}
{Eitner}, P., {Bergemann}, M., {Ruiter}, A.~J., {et~al.} 2023, \aap, 677, A151

\bibitem[{{Fink} {et~al.}(2014){Fink}, {Kromer}, {Seitenzahl}, {Ciaraldi-Schoolmann}, {R{\"o}pke}, {Sim}, {Pakmor}, {Ruiter}, \& {Hillebrandt}}]{fink:2014}
{Fink}, M., {Kromer}, M., {Seitenzahl}, I.~R., {et~al.} 2014, \mnras, 438, 1762

\bibitem[{{Foley} {et~al.}(2013){Foley}, {Challis}, {Chornock}, {Ganeshalingam}, {Li}, {Marion}, {Morrell}, {Pignata}, {Stritzinger}, {Silverman}, {Wang}, {Anderson}, {Filippenko}, {Freedman}, {Hamuy}, {Jha}, {Kirshner}, {McCully}, {Persson}, {Phillips}, {Reichart}, \& {Soderberg}}]{Foley:2013}
{Foley}, R.~J., {Challis}, P.~J., {Chornock}, R., {et~al.} 2013, \apj, 767, 57

\bibitem[{{Foley} {et~al.}(2009){Foley}, {Chornock}, {Filippenko}, {Ganeshalingam}, {Kirshner}, {Li}, {Cenko}, {Challis}, {Friedman}, {Modjaz}, {Silverman}, \& {Wood-Vasey}}]{Foley:2009}
{Foley}, R.~J., {Chornock}, R., {Filippenko}, A.~V., {et~al.} 2009, \aj, 138, 376

\bibitem[{{Grevesse} {et~al.}(2007){Grevesse}, {Asplund}, \& {Sauval}}]{Grevesse:2007}
{Grevesse}, N., {Asplund}, M., \& {Sauval}, A.~J. 2007, \ssr, 130, 105

\bibitem[{{Gronow} {et~al.}(2021{\natexlab{a}}){Gronow}, {Collins}, {Sim}, \& {R{\"o}pke}}]{gronow:21b}
{Gronow}, S., {Collins}, C.~E., {Sim}, S.~A., \& {R{\"o}pke}, F.~K. 2021{\natexlab{a}}, \aap, 649, A155

\bibitem[{{Gronow} {et~al.}(2021{\natexlab{b}}){Gronow}, {C{\^o}t{\'e}}, {Lach}, {Seitenzahl}, {Collins}, {Sim}, \& {R{\"o}pke}}]{gronow:21gce}
{Gronow}, S., {C{\^o}t{\'e}}, B., {Lach}, F., {et~al.} 2021{\natexlab{b}}, \aap, 656, A94

\bibitem[{{Hartmann} {et~al.}(1985){Hartmann}, {Woosley}, \& {El Eid}}]{Hartman:1985}
{Hartmann}, D., {Woosley}, S.~E., \& {El Eid}, M.~F. 1985, \apj, 297, 837

\bibitem[{{Hillebrandt} {et~al.}(2013){Hillebrandt}, {Kromer}, {R{\"o}pke}, \& {Ruiter}}]{hillebrandt:2013}
{Hillebrandt}, W., {Kromer}, M., {R{\"o}pke}, F.~K., \& {Ruiter}, A.~J. 2013, Frontiers of Physics, 8, 116

\bibitem[{{Hughes} {et~al.}(2008){Hughes}, {Gibson}, {Carigi}, {S{\'a}nchez-Bl{\'a}zquez}, {Chavez}, \& {Lambert}}]{Hughes:2008}
{Hughes}, G.~L., {Gibson}, B.~K., {Carigi}, L., {et~al.} 2008, \mnras, 390, 1710

\bibitem[{{Iben} \& {Tutukov}(1984)}]{iben:1984}
{Iben}, I., J. \& {Tutukov}, A.~V. 1984, \apjs, 54, 335

\bibitem[{{Iwamoto} {et~al.}(1999){Iwamoto}, {Brachwitz}, {Nomoto}, {Kishimoto}, {Umeda}, {Hix}, \& {Thielemann}}]{iwamoto:99}
{Iwamoto}, K., {Brachwitz}, F., {Nomoto}, K., {et~al.} 1999, \apjs, 125, 439

\bibitem[{{Jha}(2017)}]{Jha:2017}
{Jha}, S.~W. 2017, in Handbook of Supernovae, ed. A.~W. {Alsabti} \& P.~{Murdin}, 375

\bibitem[{{Johansson} {et~al.}(2016){Johansson}, {Woods}, {Gilfanov}, {Sarzi}, {Chen}, \& {Oh}}]{Johansson:2016}
{Johansson}, J., {Woods}, T.~E., {Gilfanov}, M., {et~al.} 2016, \mnras, 461, 4505

\bibitem[{{Karakas} \& {Lattanzio}(2014)}]{karakas:14}
{Karakas}, A.~I. \& {Lattanzio}, J.~C. 2014, \pasa, 31, e030

\bibitem[{{Karambelkar} {et~al.}(2022){Karambelkar}, {Kasliwal}, {Maguire}, {Anand}, {Andreoni}, {De}, {Drake}, {Duev}, {Graham}, {Kool}, {Laher}, {Magee}, {Mahabal}, {Medford}, {Perley}, {Rigault}, {Rusholme}, {Schulze}, {Sharma}, {Sollerman}, {Tzanidakis}, {Walters}, \& {Yao}}]{Karambelkar:2022}
{Karambelkar}, V., {Kasliwal}, M., {Maguire}, K., {et~al.} 2022, in American Astronomical Society Meeting Abstracts, Vol.~54, American Astronomical Society Meeting Abstracts, 317.07

\bibitem[{{Keegans} {et~al.}(2023){Keegans}, {Pignatari}, {Stancliffe}, {Travaglio}, {Jones}, {Gibson}, {Townsley}, {Miles}, {Shen}, \& {Few}}]{Keegans:2023}
{Keegans}, J.~D., {Pignatari}, M., {Stancliffe}, R.~J., {et~al.} 2023, \apjs, 268, 8

\bibitem[{{Kirby} {et~al.}(2019){Kirby}, {Xie}, {Guo}, {de los Reyes}, {Bergemann}, {Kovalev}, {Shen}, {Piro}, \& {McWilliam}}]{kirby:2019}
{Kirby}, E.~N., {Xie}, J.~L., {Guo}, R., {et~al.} 2019, \apj, 881, 45

\bibitem[{{Kobayashi} {et~al.}(2020{\natexlab{a}}){Kobayashi}, {Karakas}, \& {Lugaro}}]{Kobayashi:2020a}
{Kobayashi}, C., {Karakas}, A.~I., \& {Lugaro}, M. 2020{\natexlab{a}}, \apj, 900, 179

\bibitem[{{Kobayashi} {et~al.}(2020{\natexlab{b}}){Kobayashi}, {Leung}, \& {Nomoto}}]{kobayashi:mn}
{Kobayashi}, C., {Leung}, S.-C., \& {Nomoto}, K. 2020{\natexlab{b}}, \apj, 895, 138

\bibitem[{{Kobayashi} \& {Nakasato}(2011)}]{Kobayashi:2011}
{Kobayashi}, C. \& {Nakasato}, N. 2011, \apj, 729, 16

\bibitem[{{Kobayashi} {et~al.}(2015){Kobayashi}, {Nomoto}, \& {Hachisu}}]{kobayashi:2015}
{Kobayashi}, C., {Nomoto}, K., \& {Hachisu}, I. 2015, \apjl, 804, L24

\bibitem[{{Kobayashi} {et~al.}(2006){Kobayashi}, {Umeda}, {Nomoto}, {Tominaga}, \& {Ohkubo}}]{kobayashi:2006}
{Kobayashi}, C., {Umeda}, H., {Nomoto}, K., {Tominaga}, N., \& {Ohkubo}, T. 2006, \apj, 653, 1145

\bibitem[{{Kromer} {et~al.}(2015){Kromer}, {Ohlmann}, {Pakmor}, {Ruiter}, {Hillebrandt}, {Marquardt}, {R{\"o}pke}, {Seitenzahl}, {Sim}, \& {Taubenberger}}]{kromer:15}
{Kromer}, M., {Ohlmann}, S.~T., {Pakmor}, R., {et~al.} 2015, \mnras, 450, 3045

\bibitem[{{Kroupa} {et~al.}(1993){Kroupa}, {Tout}, \& {Gilmore}}]{kroupa:1993}
{Kroupa}, P., {Tout}, C.~A., \& {Gilmore}, G. 1993, \mnras, 262, 545

\bibitem[{{Kubryk} {et~al.}(2015){Kubryk}, {Prantzos}, \& {Athanassoula}}]{kubryk:2015}
{Kubryk}, M., {Prantzos}, N., \& {Athanassoula}, E. 2015, \aap, 580, A126

\bibitem[{{Lach} {et~al.}(2020){Lach}, {R{\"o}pke}, {Seitenzahl}, {Cot{\'e}}, {Gronow}, \& {Ruiter}}]{lach:2020}
{Lach}, F., {R{\"o}pke}, F.~K., {Seitenzahl}, I.~R., {et~al.} 2020, \aap, 644, A118

\bibitem[{{Leung} \& {Nomoto}(2018)}]{leung2018}
{Leung}, S.-C. \& {Nomoto}, K. 2018, \apj, 861, 143

\bibitem[{{Leung} \& {Nomoto}(2020)}]{leung:20}
{Leung}, S.-C. \& {Nomoto}, K. 2020, \apj, 888, 80

\bibitem[{{Leung} {et~al.}(2023){Leung}, {Nomoto}, \& {Suzuki}}]{Leung:2023}
{Leung}, S.-C., {Nomoto}, K., \& {Suzuki}, T. 2023, \apj, 948, 80

\bibitem[{{Limongi} \& {Chieffi}(2018)}]{limongi2018}
{Limongi}, M. \& {Chieffi}, A. 2018, \apjs, 237, 13

\bibitem[{{Liu} \& {Wang}(2020)}]{liu:2020}
{Liu}, D. \& {Wang}, B. 2020, \mnras, 494, 3422

\bibitem[{{Liu} {et~al.}(2015){Liu}, {Moriya}, {Stancliffe}, \& {Wang}}]{liu:2015}
{Liu}, Z.-W., {Moriya}, T.~J., {Stancliffe}, R.~J., \& {Wang}, B. 2015, \aap, 574, A12

\bibitem[{{Livio} \& {Mazzali}(2018)}]{mario:2018}
{Livio}, M. \& {Mazzali}, P. 2018, \physrep, 736, 1

\bibitem[{{Livne} \& {Arnett}(1995)}]{Livne:1995}
{Livne}, E. \& {Arnett}, D. 1995, \apj, 452, 62

\bibitem[{{Lomaeva} {et~al.}(2019){Lomaeva}, {J{\"o}nsson}, {Ryde}, {Schultheis}, \& {Thorsbro}}]{Lomaeva:2019}
{Lomaeva}, M., {J{\"o}nsson}, H., {Ryde}, N., {Schultheis}, M., \& {Thorsbro}, B. 2019, \aap, 625, A141

\bibitem[{{Mallinson} {et~al.}(2024){Mallinson}, {Lind}, {Amarsi}, \& {Youakim}}]{Mallinson:2024}
{Mallinson}, J.~W.~E., {Lind}, K., {Amarsi}, A.~M., \& {Youakim}, K. 2024, \aap, 687, A5

\bibitem[{{Mannucci} {et~al.}(2006){Mannucci}, {Della Valle}, \& {Panagia}}]{man:2006}
{Mannucci}, F., {Della Valle}, M., \& {Panagia}, N. 2006, \mnras, 370, 773

\bibitem[{Maoz \& Mannucci(2012)}]{maoz_mannucci_2012}
Maoz, D. \& Mannucci, F. 2012, Publications of the Astronomical Society of Australia, 29, 447–465

\bibitem[{{Maoz} {et~al.}(2012){Maoz}, {Mannucci}, \& {Brandt}}]{maoz:12}
{Maoz}, D., {Mannucci}, F., \& {Brandt}, T.~D. 2012, \mnras, 426, 3282

\bibitem[{{Matteucci}(2021)}]{matteucci:2021}
{Matteucci}, F. 2021, \aapr, 29, 5

\bibitem[{{Matteucci} \& {Greggio}(1986)}]{matteucci:1986}
{Matteucci}, F. \& {Greggio}, L. 1986, \aap, 154, 279

\bibitem[{{Matteucci} {et~al.}(2009){Matteucci}, {Spitoni}, {Recchi}, \& {Valiante}}]{matteucci:2009}
{Matteucci}, F., {Spitoni}, E., {Recchi}, S., \& {Valiante}, R. 2009, \aap, 501, 531

\bibitem[{{Minchev} {et~al.}(2013){Minchev}, {Chiappini}, \& {Martig}}]{minchev2013}
{Minchev}, I., {Chiappini}, C., \& {Martig}, M. 2013, \aap, 558, A9

\bibitem[{{Mishenina} {et~al.}(2017){Mishenina}, {Pignatari}, {C{\^o}t{\'e}}, {Thielemann}, {Soubiran}, {Basak}, {Gorbaneva}, {Korotin}, {Kovtyukh}, {Wehmeyer}, {Bisterzo}, {Travaglio}, {Gibson}, {Jordan}, {Paul}, {Ritter}, {Herwig}, \& {NuGrid Collaboration}}]{Mishenina:2017}
{Mishenina}, T., {Pignatari}, M., {C{\^o}t{\'e}}, B., {et~al.} 2017, \mnras, 469, 4378

\bibitem[{{Nomoto} {et~al.}(2013){Nomoto}, {Kobayashi}, \& {Tominaga}}]{nomoto13}
{Nomoto}, K., {Kobayashi}, C., \& {Tominaga}, N. 2013, \araa, 51, 457

\bibitem[{{Nomoto} {et~al.}(1984){Nomoto}, {Thielemann}, \& {Yokoi}}]{Nomoto:1984}
{Nomoto}, K., {Thielemann}, F.~K., \& {Yokoi}, K. 1984, \apj, 286, 644

\bibitem[{{Nomoto} {et~al.}(2006){Nomoto}, {Tominaga}, {Umeda}, {Kobayashi}, \& {Maeda}}]{Nomoto:2006}
{Nomoto}, K., {Tominaga}, N., {Umeda}, H., {Kobayashi}, C., \& {Maeda}, K. 2006, \nphysa, 777, 424

\bibitem[{{Pagel}(2009)}]{pagel:2009}
{Pagel}, B. E.~J. 2009, {Nucleosynthesis and Chemical Evolution of Galaxies}

\bibitem[{{Pakmor} {et~al.}(2011){Pakmor}, {Hachinger}, {R{\"o}pke}, \& {Hillebrandt}}]{pakmor:2011}
{Pakmor}, R., {Hachinger}, S., {R{\"o}pke}, F.~K., \& {Hillebrandt}, W. 2011, \aap, 528, A117

\bibitem[{{Pakmor} {et~al.}(2012){Pakmor}, {Kromer}, {Taubenberger}, {Sim}, {R{\"o}pke}, \& {Hillebrandt}}]{pakmor:12}
{Pakmor}, R., {Kromer}, M., {Taubenberger}, S., {et~al.} 2012, \apjl, 747, L10

\bibitem[{{Palla}(2021)}]{palla:2021}
{Palla}, M. 2021, \mnras, 503, 3216

\bibitem[{{Perlmutter} {et~al.}(1999){Perlmutter}, {Aldering}, {Goldhaber}, {Knop}, {Nugent}, {Castro}, {Deustua}, {Fabbro}, {Goobar}, {Groom}, {Hook}, {Kim}, {Kim}, {Lee}, {Nunes}, {Pain}, {Pennypacker}, {Quimby}, {Lidman}, {Ellis}, {Irwin}, {McMahon}, {Ruiz-Lapuente}, {Walton}, {Schaefer}, {Boyle}, {Filippenko}, {Matheson}, {Fruchter}, {Panagia}, {Newberg}, {Couch}, \& {Project}}]{perlmutter:1999}
{Perlmutter}, S., {Aldering}, G., {Goldhaber}, G., {et~al.} 1999, \apj, 517, 565

\bibitem[{{Phillips}(1993)}]{phillips:93}
{Phillips}, M.~M. 1993, \apjl, 413, L105

\bibitem[{{Prantzos}(2008)}]{prantzos:2008}
{Prantzos}, N. 2008, in EAS Publications Series, Vol.~32, EAS Publications Series, ed. C.~{Charbonnel} \& J.~P. {Zahn}, 311--356

\bibitem[{{Prantzos} {et~al.}(2018){Prantzos}, {Abia}, {Limongi}, {Chieffi}, \& {Cristallo}}]{prantzos:18}
{Prantzos}, N., {Abia}, C., {Limongi}, M., {Chieffi}, A., \& {Cristallo}, S. 2018, \mnras, 476, 3432

\bibitem[{{Riess} {et~al.}(1998){Riess}, {Filippenko}, {Challis}, {Clocchiatti}, {Diercks}, {Garnavich}, {Gilliland}, {Hogan}, {Jha}, {Kirshner}, {Leibundgut}, {Phillips}, {Reiss}, {Schmidt}, {Schommer}, {Smith}, {Spyromilio}, {Stubbs}, {Suntzeff}, \& {Tonry}}]{riess:1998}
{Riess}, A.~G., {Filippenko}, A.~V., {Challis}, P., {et~al.} 1998, \aj, 116, 1009

\bibitem[{{Ritter} {et~al.}(2018){Ritter}, {C{\^o}t{\'e}}, {Herwig}, {Navarro}, \& {Fryer}}]{rit18}
{Ritter}, C., {C{\^o}t{\'e}}, B., {Herwig}, F., {Navarro}, J.~F., \& {Fryer}, C.~L. 2018, \apjs, 237, 42

\bibitem[{{Romano} {et~al.}(2010){Romano}, {Karakas}, {Tosi}, \& {Matteucci}}]{Romano:2010}
{Romano}, D., {Karakas}, A.~I., {Tosi}, M., \& {Matteucci}, F. 2010, \aap, 522, A32

\bibitem[{{R{\"o}pke}(2007)}]{Ropke:2007}
{R{\"o}pke}, F.~K. 2007, \apj, 668, 1103

\bibitem[{{R{\"o}pke} {et~al.}(2007){R{\"o}pke}, {Hillebrandt}, {Schmidt}, {Niemeyer}, {Blinnikov}, \& {Mazzali}}]{Ropke:2007b}
{R{\"o}pke}, F.~K., {Hillebrandt}, W., {Schmidt}, W., {et~al.} 2007, \apj, 668, 1132

\bibitem[{{Ruiter}(2020)}]{Ruiter:2021}
{Ruiter}, A.~J. 2020, IAU Symposium, 357, 1

\bibitem[{{Ruiter} {et~al.}(2009){Ruiter}, {Belczynski}, \& {Fryer}}]{ruiter:09}
{Ruiter}, A.~J., {Belczynski}, K., \& {Fryer}, C. 2009, \apj, 699, 2026

\bibitem[{{Ruiter} {et~al.}(2011){Ruiter}, {Belczynski}, {Sim}, {Hillebrandt}, {Fryer}, {Fink}, \& {Kromer}}]{Ruiter:2011}
{Ruiter}, A.~J., {Belczynski}, K., {Sim}, S.~A., {et~al.} 2011, \mnras, 417, 408

\bibitem[{{Ruiter} {et~al.}(2014){Ruiter}, {Belczynski}, {Sim}, {Seitenzahl}, \& {Kwiatkowski}}]{ruiter:14}
{Ruiter}, A.~J., {Belczynski}, K., {Sim}, S.~A., {Seitenzahl}, I.~R., \& {Kwiatkowski}, D. 2014, \mnras, 440, L101

\bibitem[{{Sanders} {et~al.}(2021){Sanders}, {Belokurov}, \& {Man}}]{Sanders:2021}
{Sanders}, J.~L., {Belokurov}, V., \& {Man}, K. T.~F. 2021, \mnras, 506, 4321

\bibitem[{{Sato} {et~al.}(2016){Sato}, {Nakasato}, {Tanikawa}, {Nomoto}, {Maeda}, \& {Hachisu}}]{sato:2016}
{Sato}, Y., {Nakasato}, N., {Tanikawa}, A., {et~al.} 2016, \apj, 821, 67

\bibitem[{{Scannapieco} \& {Bildsten}(2005)}]{scannapieco:05}
{Scannapieco}, E. \& {Bildsten}, L. 2005, \apjl, 629, L85

\bibitem[{{Seitenzahl} {et~al.}(2013{\natexlab{a}}){Seitenzahl}, {Cescutti}, {R{\"o}pke}, {Ruiter}, \& {Pakmor}}]{seitenzahl:Mn}
{Seitenzahl}, I.~R., {Cescutti}, G., {R{\"o}pke}, F.~K., {Ruiter}, A.~J., \& {Pakmor}, R. 2013{\natexlab{a}}, \aap, 559, L5

\bibitem[{{Seitenzahl} {et~al.}(2013{\natexlab{b}}){Seitenzahl}, {Ciaraldi-Schoolmann}, {R{\"o}pke}, {Fink}, {Hillebrandt}, {Kromer}, {Pakmor}, {Ruiter}, {Sim}, \& {Taubenberger}}]{seitenzahl:12}
{Seitenzahl}, I.~R., {Ciaraldi-Schoolmann}, F., {R{\"o}pke}, F.~K., {et~al.} 2013{\natexlab{b}}, \mnras, 429, 1156

\bibitem[{{Seitenzahl} \& {Townsley}(2017)}]{Seitenzahl:2017}
{Seitenzahl}, I.~R. \& {Townsley}, D.~M. 2017, in Handbook of Supernovae, ed. A.~W. {Alsabti} \& P.~{Murdin}, 1955

\bibitem[{{Shen} {et~al.}(2018){Shen}, {Kasen}, {Miles}, \& {Townsley}}]{shen:2018}
{Shen}, K.~J., {Kasen}, D., {Miles}, B.~J., \& {Townsley}, D.~M. 2018, \apj, 854, 52

\bibitem[{{Shen} \& {Moore}(2014)}]{shen:2014}
{Shen}, K.~J. \& {Moore}, K. 2014, \apj, 797, 46

\bibitem[{{Sim} {et~al.}(2010){Sim}, {R{\"o}pke}, {Hillebrandt}, {Kromer}, {Pakmor}, {Fink}, {Ruiter}, \& {Seitenzahl}}]{sim:2010}
{Sim}, S.~A., {R{\"o}pke}, F.~K., {Hillebrandt}, W., {et~al.} 2010, \apjl, 714, L52

\bibitem[{{Sitnova} {et~al.}(2022){Sitnova}, {Yakovleva}, {Belyaev}, \& {Mashonkina}}]{Sitnova:2022}
{Sitnova}, T.~M., {Yakovleva}, S.~A., {Belyaev}, A.~K., \& {Mashonkina}, L.~I. 2022, \mnras, 515, 1510

\bibitem[{{Snaith} {et~al.}(2015){Snaith}, {Haywood}, {Di Matteo}, {Lehnert}, {Combes}, {Katz}, \& {G{\'o}mez}}]{snaith:2015}
{Snaith}, O., {Haywood}, M., {Di Matteo}, P., {et~al.} 2015, \aap, 578, A87

\bibitem[{{Soker}(2019)}]{Soker:2019}
{Soker}, N. 2019, \nar, 87, 101535

\bibitem[{{Srivastav} {et~al.}(2022){Srivastav}, {Smartt}, {Huber}, {Chambers}, {Angus}, {Chen}, {Callan}, {Gillanders}, {McBrien}, {Sim}, {Fulton}, {Hjorth}, {Smith}, {Young}, {Auchettl}, {Anderson}, {Pignata}, {de Boer}, {Lin}, \& {Magnier}}]{srivastav:2022}
{Srivastav}, S., {Smartt}, S.~J., {Huber}, M.~E., {et~al.} 2022, \mnras, 511, 2708

\bibitem[{{Steinmetz} {et~al.}(1992){Steinmetz}, {Muller}, \& {Hillebrandt}}]{Steinmetz:1992}
{Steinmetz}, M., {Muller}, E., \& {Hillebrandt}, W. 1992, \aap, 254, 177

\bibitem[{{Taubenberger}(2017)}]{tauben:2017}
{Taubenberger}, S. 2017, in Handbook of Supernovae, ed. A.~W. {Alsabti} \& P.~{Murdin}, 317

\bibitem[{{Temmink} {et~al.}(2020){Temmink}, {Toonen}, {Zapartas}, {Justham}, \& {G{\"a}nsicke}}]{Temmink:2020}
{Temmink}, K.~D., {Toonen}, S., {Zapartas}, E., {Justham}, S., \& {G{\"a}nsicke}, B.~T. 2020, \aap, 636, A31

\bibitem[{{Thielemann} {et~al.}(2003){Thielemann}, {Argast}, {Brachwitz}, {Hix}, {H{\"o}flich}, {Liebend{\"o}rfer}, {Martinez-Pinedo}, {Mezzacappa}, {Nomoto}, \& {Panov}}]{Thielemann:2003}
{Thielemann}, F.~K., {Argast}, D., {Brachwitz}, F., {et~al.} 2003, in From Twilight to Highlight: The Physics of Supernovae, ed. W.~{Hillebrandt} \& B.~{Leibundgut}, 331

\bibitem[{{Thielemann} {et~al.}(1996){Thielemann}, {Nomoto}, \& {Hashimoto}}]{Thielemann:1996}
{Thielemann}, F.-K., {Nomoto}, K., \& {Hashimoto}, M.-A. 1996, \apj, 460, 408

\bibitem[{{Timmes} {et~al.}(1995){Timmes}, {Woosley}, \& {Weaver}}]{timmes:1995}
{Timmes}, F.~X., {Woosley}, S.~E., \& {Weaver}, T.~A. 1995, \apjs, 98, 617

\bibitem[{{Toonen} {et~al.}(2014){Toonen}, {Claeys}, {Mennekens}, \& {Ruiter}}]{toonen:2014}
{Toonen}, S., {Claeys}, J.~S.~W., {Mennekens}, N., \& {Ruiter}, A.~J. 2014, \aap, 562, A14

\bibitem[{{Totani} {et~al.}(2008){Totani}, {Morokuma}, {Oda}, {Doi}, \& {Yasuda}}]{Totani:2008}
{Totani}, T., {Morokuma}, T., {Oda}, T., {Doi}, M., \& {Yasuda}, N. 2008, \pasj, 60, 1327

\bibitem[{{Townsley} {et~al.}(2016){Townsley}, {Miles}, {Timmes}, {Calder}, \& {Brown}}]{townsley:16}
{Townsley}, D.~M., {Miles}, B.~J., {Timmes}, F.~X., {Calder}, A.~C., \& {Brown}, E.~F. 2016, \apjs, 225, 3

\bibitem[{{Valenti} {et~al.}(2009){Valenti}, {Pastorello}, {Cappellaro}, {Benetti}, {Mazzali}, {Manteca}, {Taubenberger}, {Elias-Rosa}, {Ferrando}, {Harutyunyan}, {Hentunen}, {Nissinen}, {Pian}, {Turatto}, {Zampieri}, \& {Smartt}}]{Valenti:2009}
{Valenti}, S., {Pastorello}, A., {Cappellaro}, E., {et~al.} 2009, \nat, 459, 674

\bibitem[{{Whelan} \& {Iben}(1973)}]{whelan:1973}
{Whelan}, J. \& {Iben}, Icko, J. 1973, \apj, 186, 1007

\bibitem[{{Woods} \& {Gilfanov}(2013)}]{Woods:2013}
{Woods}, T.~E. \& {Gilfanov}, M. 2013, \mnras, 432, 1640

\bibitem[{{Woosley} \& {Weaver}(1994)}]{woosley:1994}
{Woosley}, S.~E. \& {Weaver}, T.~A. 1994, \apj, 423, 371

\bibitem[{{Woosley} \& {Weaver}(1995)}]{Woosley:1995}
{Woosley}, S.~E. \& {Weaver}, T.~A. 1995, \apjs, 101, 181

\bibitem[{{Zhao} {et~al.}(2016){Zhao}, {Mashonkina}, {Yan}, {Alexeeva}, {Kobayashi}, {Pakhomov}, {Shi}, {Sitnova}, {Tan}, {Zhang}, {Zhang}, {Zhou}, {Bolte}, {Chen}, {Li}, {Liu}, \& {Zhai}}]{Zhao:2016}
{Zhao}, G., {Mashonkina}, L., {Yan}, H.~L., {et~al.} 2016, \apj, 833, 225

\end{thebibliography}
\bibliographystyle{aa}

\begin{appendix}

\onecolumn{}
\section{Galactic chemical evolution models with SN Ia yield combinations} \label{Appendix}

\begin{figure*}[h!]
    \centering
    \includegraphics[width=1.0\linewidth]{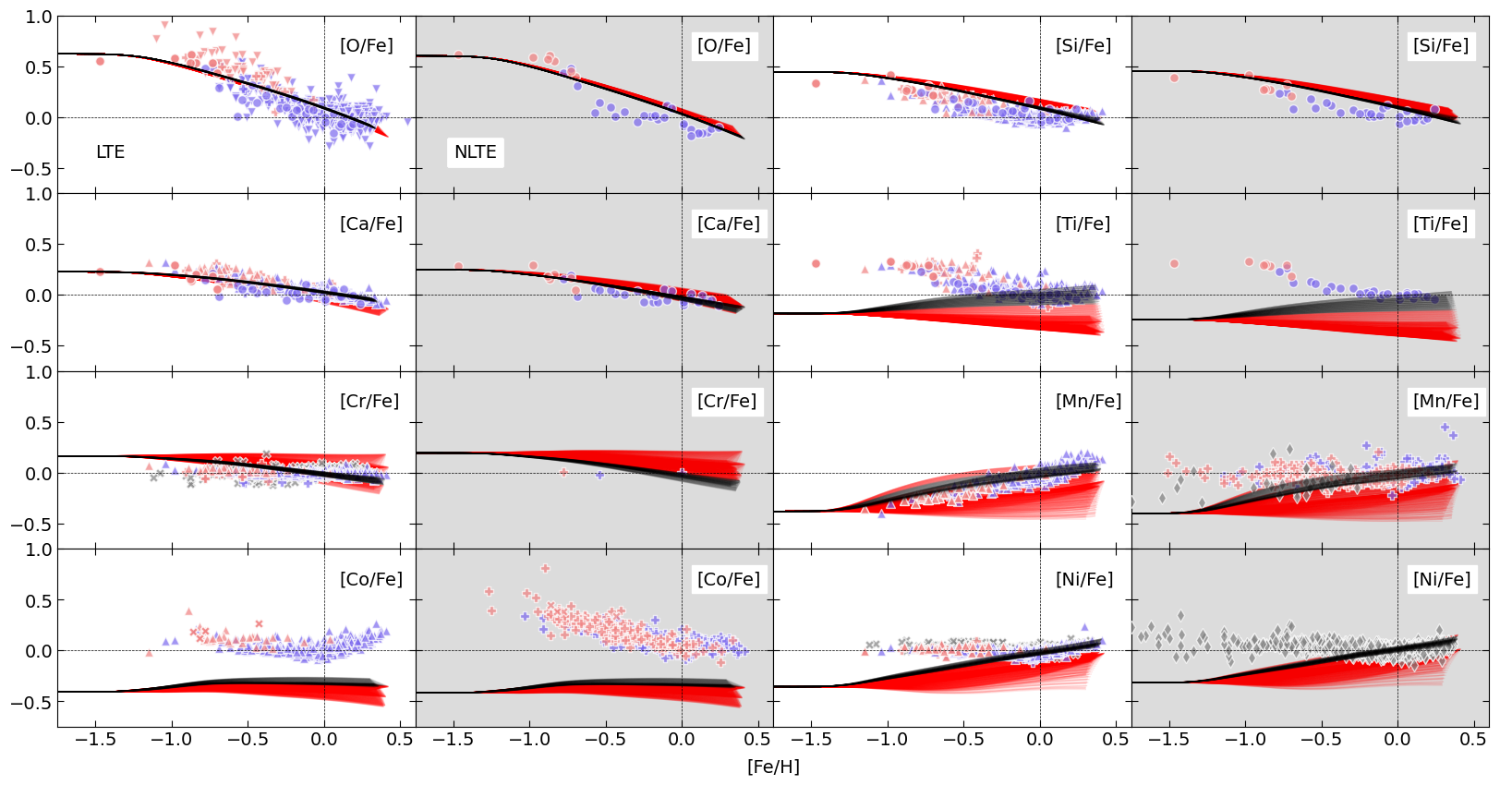}
    \caption{GCE tracks for N13 CCSN yields with different SN Ia yield combinations (see Section \ref{sec:combos} for details) with no contribution from SNe Iax. The GCE tracks with black lines indicate the models that have a $\chi^2$ score in 84th percentile. Observational data are the same as in Fig. \ref{fig:N13_GCE} \& \ref{fig:LC18_GCE}.}
    \label{fig:N13_GCE_chi}
\end{figure*}

\begin{figure*}[h!]
    \centering
    \includegraphics[width=1.0\linewidth]{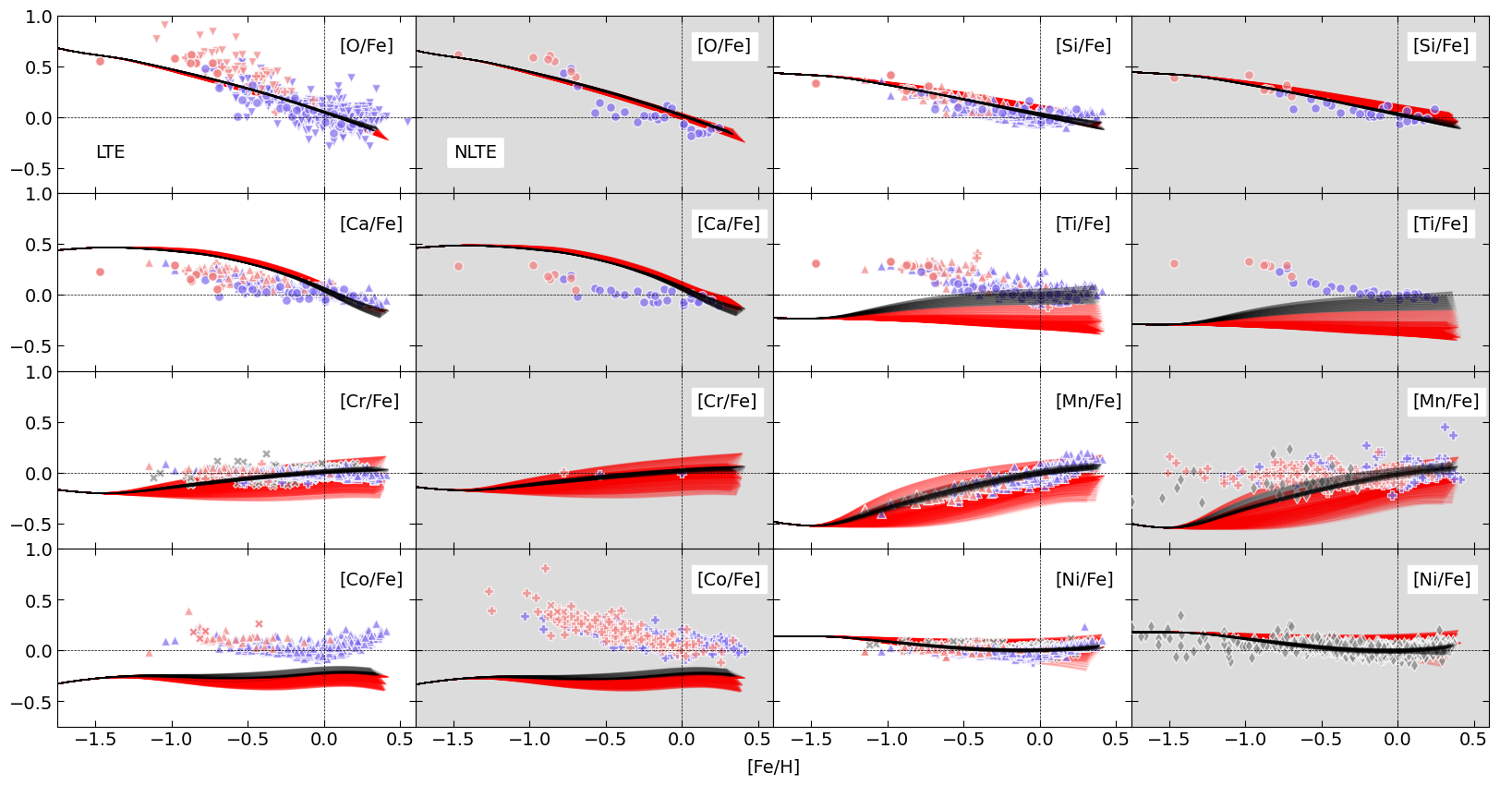}
    \caption{Same as Fig. \ref{fig:N13_GCE_chi}, but for LC18 CCSN yields.}
    \label{fig:LC18_GCE_chi}
\end{figure*}

Fig. \ref{fig:N13_GCE_chi} \& \ref{fig:LC18_GCE_chi} show the predicted chemical evolution of intermediate mass $\alpha$-elements and Fe-peak elements for each of the 1140 GCE models calculated using a combination of the sub- and near-$M_{\text{Ch}}$ yields investigated in this work (except for PDf explosions). Each GCE track is colour coded based on the goodness of fit to the observational data according to the $\chi^2$ test described by Equation \ref{eq:chi}.

\end{appendix}

\end{document}